\documentclass[preprint,authoryear,12pt]{elsarticle}
\usepackage{amsmath, mathtools, color}
\usepackage{float}

\usepackage{accents}
\usepackage{stackengine}

\usepackage{graphicx, geometry,graphicx}
\usepackage{epstopdf}
        \usepackage{dutchcal}

\newcommand{\be}{\begin{equation}}
\newcommand{\en}{\end{equation}}

\renewcommand{\vec}[1]{\boldsymbol{#1}}

\def \curl{\mbox{curl\hskip 1pt}}

\def \div{\mbox{div\hskip 1pt}}

\def \tr{\mbox{tr\hskip 1pt}}
\def \grad{\mbox{grad\hskip 1pt}}

\begin{document}

\begin{frontmatter}

\title{Finite bending and pattern evolution of the associated instability for a dielectric elastomer slab}

\author[1,2]{Yipin Su}
\author[2]{Bin Wu}
\author[2]{Weiqiu Chen}
\author[1,2]{Michel Destrade}

\address[1]{School of Mathematics, Statistics and Applied Mathematics, \\
NUI Galway, University Road, Galway, Ireland\\[12pt]}
\address[2]{Department of Engineering Mechanics,\\
Zhejiang University, Hangzhou 310027, P.R. China}

%
%

\begin{abstract}
We investigate the finite bending and the associated bending instability of an incompressible dielectric slab subject to a combination of applied voltage and axial compression, using nonlinear electro-elasticity theory and its incremental version. We first study the static finite bending deformation of the slab. We then derive the three-dimensional equations for the onset of small-amplitude wrinkles superimposed upon the finite bending. We use the surface impedance matrix method to build a robust numerical procedure for solving the resulting dispersion equations and determining the wrinkled shape of the slab at the onset of buckling. Our analysis is valid for dielectrics modeled by a general free energy function. We then present illustrative numerical calculations for ideal neo-Hookean dielectrics. In that case, we provide an explicit treatment of the boundary value problem of the finite bending and derive closed-form expressions for the stresses and electric field in the body. For the incremental deformations, we validate our analysis by recovering existing results in more specialized contexts. We show that the applied voltage has a destabilizing effect on the bending instability of the slab, while the effect of the axial load is more complex: when the voltage is applied, changing the axial loading will influence the true electric field in the body, and induce competitive effects between the circumferential instability due to the voltage and the axial instability due to the axial compression. We even find circumstances where both instabilities cohabit to create two-dimensional patterns on the inner face of the bent sector.
\end{abstract}

\begin{keyword}
finite bending \sep bending instability \sep surface impedance matrix method \sep two-dimensional wrinkles
\end{keyword}

\end{frontmatter}


\section{Introduction}
An elastic rectangular slab can be bent into a cylindrical sector under the application of moments on the lateral faces, and the bending angle depends on the applied moments, the dimensions and the material properties of the slab. The finite bending deformation of incompressible soft materials is well captured by the theory of nonlinear elasticity \citep{Rivlin49, Green54, Truesdell60, Ogden97}. Generally speaking, the inner surface of a bent slab is contracted circumferentially, and the outer surface is stretched. Experimental observations indicate that wrinkles and creases will appear on the compressed surface of a bent rubber slab if the circumferential stretch of the inner surface reaches a critical value, i.e., the so-called bending instability occurs \citep{Gent99}. This phenomenon can be predicted by the theory of incremental nonlinear elasticity \citep{Triantafyllidis80, Gilchrist09, Destrade09, Roccabianca10, Destrade14}.

Dielectric elastomers are novel smart materials with the ability to convert mechanical energy into electrical energy, and vice versa. Dielectric elastomers have attracted considerable attention from academia and industry alike because, compared with other smart materials like electroactive ceramics and shape memory alloys, they have the advantages of fast response, high-sensitivity, low noise and large actuation strain, making them ideal candidates to develop high-performance devices such as actuators, soft robots, artificial muscles, phononic devices and energy harvesters \citep{Bar-Cohen04, Kim07, Rasmussen12, Brochu10, Galich17, Getz17, Wu18}. Generally, a dielectric actuator is composed of a soft elastomeric material sandwiched between two compliant electrodes (typically, by brushing on carbon grease). Application of a voltage across the thickness of the actuator generates electrostatic forces, which lead to a reduction in the thickness and an expansion in the area of the actuator. Based on this mechanism, various dielectric devices have been designed to achieve giant actuation strains \citep{	Pelrine00, O'Halloran08, Zhang17}.

To understand the electromechanical coupling effect and predict the nonlinear response of dielectric elastomers subject to electromechanical loadings, a nonlinear field theory is required. Arguably, \cite{Toupin56} was the first to develop a general nonlinear theory of electro-elasticity. Much effort has been devoted to the development of this theory in the last two decades \citep{Ericksen07, Suo08, Liu13, Dorf16}, driven by recent applications in the real-world. So far, several finite deformations of dielectric structures have been investigated theoretically, including simple shear of a dielectric slab \citep{Dorf05}, in-plane homogeneous deformation of a dielectric plate \citep{Dorf14a}, extension and inflation of a dielectric tube \citep{Dorf06, Zhu10} and a multilayer dielectric tube \citep{Bortot18}, inflation of a dielectric sphere \citep{Li13, Dorf14b} and of a multilayer dielectric sphere \citep{Bortot17}.

\begin{figure}[h!]
\centering
\includegraphics[width=0.8\textwidth]{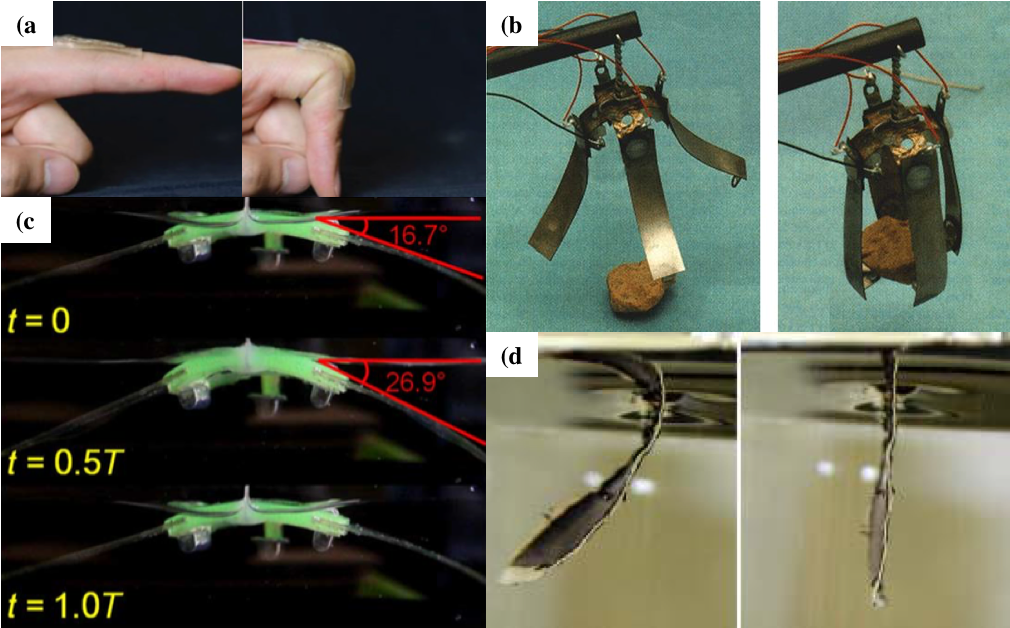}
\caption{
{\footnotesize
Bending deformations in dielectric devices: (a) A strain sensor consisting of a stretchable dielectric sandwiched between two flexible ionic conductors attached to a straight finger: the bending of the finger stretches the sensor \citep{Sun14}; (b) A 4-finger dielectric gripper: this actuator induces voltage-driven bending to lift a rock \citep{Bar-Cohen02}; (c) Bending variations of the soft body and fins of a soft electronic ``fish'' made of dielectric elastomer and ionically conductive hydrogel: the fish can swim at a fast speed driven by periodical bending deformations \citep{Li17}; (d) A dielectric actuator with a significant voltage-driven bending response \citep{Bar-Cohen02}.
}
}
\label{figure1}
\end{figure}

Finite bending deformation is common in devices based on dielectric elastomers, see examples in Figure \ref{figure1}, but little attention has been devoted to the theoretical analysis of this deformation for dielectric structures. 
\cite{Wissman14} studied the pure bending of a dielectric elastomer actuator which contains inextensible but flexible frames. 
They simplified the kinematics by assuming plane strain deformation and modeled the bending deformation using elastic shell theory based on the principle of minimum potential energy. 
Good agreement between theoretical and experimental results was achieved for a neo-Hookean constitutive law, but the prediction is valid only for small strain deformation. 
\cite{Li14} investigated the bending deformation of a dielectric spring-roll. 
The allowable bending of the actuator was determined by considering several failure models, including electromechanical instability, electrical breakdown, and tensile rupture. 
There also, the small strain assumption was adopted to simplify the problem. 
Only recently was a theoretical study on the finite bending of a dielectric actuator based on the three-dimensional nonlinear electro-elasticity made available \citep{He17}. 
There, the authors considered an actuator consisting of a hyperelastic layer and two pre-stretched dielectric elastomer layers, which bends once a voltage is applied through the thickness of the dielectric layer. 
That analysis was concerned with static finite bending under the plane strain assumption but not with the associated bending instability. 

In this paper, we propose a theoretical analysis of finite bending deformation and the associated bending instability of an incompressible dielectric slab subject to the combined action of electrical voltage and mechanical loads. We focus on how finite bending and bending instability of a dielectric slab are influenced by tuning the applied voltage, the structural parameters and the axial compression. 

The paper is structured as follows. 
In Section \ref{section2}, we briefly recall the general equations of the nonlinear theory of electro-elasticity and the associated linear incremental field theory \citep{Dorf16}. 
We then specialize the general theory to the problems of the finite bending and the linearized incremental motion superposed upon the bending of a dielectric slab modeled by any form of energy function (Section \ref{section3}). 
We arrange the governing incremental equations in the Stroh form and then use the surface impedance matrix method to obtain a robust numerical procedure for deriving the bending and compression thresholds for the onset of the instability. We find the corresponding wrinkled shape of the slab when buckling occurs. 
In Section \ref{section4}, we present numerical calculations for an ideal neo-Hookean dielectric slab to elucidate the influence of the applied voltage, of the structural parameters and of the axial compression on the finite bending and the associated buckling behavior. 
We show analytically that only moments are required to drive the large bending of the slab. 
We find that both the applied voltage and the axial constraint pose a destabilizing effect on the slab, while these two effects compete with each other because compressing the slab will decrease the true electric field in the solid. 
We also find under which circumstances can a two-dimensional buckling pattern happen, where  circumferential and axial wrinkles co-exist. 
Finally in Section \ref{section5}, we draw some conclusions.


\section{Basic formulation}
\label{section2}


In this section we propose a brief overview of the governing equations for finite electro-elasticity and its associated incremental theory. Interested readers are referred to the textbook by \cite{Dorf16} for more detailed background on this topic.


\subsection{Finite electro-elasticity}


Consider a deformable continuous electrostatic body which, at time $t_0$, occupies an undeformed, stress-free configuration $B_r$, with boundary $\partial B_r$ and outward unit normal vector $\vec N$. Assume that the body is subject to a (true) electric field $\vec E$, with an associated (true) electric displacement $\vec D$. A material particle in $B_r$ labeled by its position vector $\vec X$ takes up the position $\vec x$ at time $t$, after a finite deformation described by the mapping $\vec x=\vec \chi(\vec X,t)$, where $\vec{\chi}$ is twice continuously differentiable. As a result, the body deforms quasi-statically into the current configuration, which is denoted by $B_t$, with the boundary $\partial B_t$ and the outward unit normal vector $\vec n$. The deformation gradient tensor is $\vec F=\partial \vec x/\partial \vec X$, with Cartesian components $F_{i\alpha}=\partial x_i/\partial X_\alpha$.  \color {black} The initial volume element $\text d\delta$ and the deformed volume element $\text d\Delta$ of the solid are related by $\text d\delta=J\text d\Delta$, where $J=\text {det} \vec F$ is the local volume ratio. \color {black}

Throughout this paper we consider incompressible dielectric elastomers, for which the internal constraint $J\equiv 1$ holds at all times. According to the theory of nonlinear electro-elasticity, by introducing an augmented free energy function $\Omega=\Omega(\vec F,\vec D_l)$, which is defined in the reference configuration, the governing equations of the body can be obtained as

\begin{equation}\label{constitive}
\vec T=\frac{\partial \Omega}{\partial \vec F}-p \vec F^{-1}, \quad \vec E_l=\frac{\partial \Omega}{\partial \vec D_l},
\end{equation}
where $\vec T=\vec F^{-1}\vec {\tau}$ is the total nominal stress, with $\vec \tau$ being the total Cauchy stress tensor, $p$ is a Lagrange multiplier associated with the incompressibility constraint, which can be determined from the boundary conditions, and the nominal electric field $\vec E_l=\vec F^{\text T}\vec E$ and the nominal electric displacement $\vec D_l=\vec F^{-1}\vec D$ are the Lagrangian counterparts of $\vec E$ and $\vec D$, respectively. The superscripts `-1' and `$\text T$' throughout this paper denote the inverse and transpose of a tensor, respectively.

Specifically, for an isotropic, incompressible, electro-elastic material, $\Omega$ can be expressed in terms of the following five invariants
 \begin{equation}\label{invariants}
{I}_1=\tr{\vec{c}}, \quad
{I}_2=\tr\left({\vec{c}^{-1}}\right),\quad
{I}_4= \vec D_l\vec\cdot \vec D_l,
\quad
{I}_5=\vec D_l\cdot\vec c\vec D_l,
\quad
{I}_6=\vec D_l \cdot\vec c^{2} \vec D_l,
\end{equation}
where $\vec c=\vec F^{\text T}\vec F$ is the right Cauchy-Green deformation tensor. Combined with Eq. \eqref{constitive}, the Cauchy stress $\vec \tau=\vec F \vec T$ and the electric field $\vec E=\vec F^{-\text T}\vec E_{l}$ are found as
\begin{equation}\label{stress}
\vec \tau=2\Omega_1\vec b+2\Omega_2\left(I_1\vec b-\vec b^2\right)-p \vec I+2\Omega_5\vec D\otimes\vec D+2\Omega_6\left(\vec D\otimes\vec b \vec D+\vec b \vec D\otimes\vec D\right),
\end{equation}
\begin{equation}\label{electric}
\vec E=2\left(\Omega_4\vec b^{-1}\vec D+\Omega_5 \vec D+\Omega_6 \vec b \vec D\right),
\end{equation}
where $\vec I$ is identity tensor, $\vec b=\vec F\vec F^\text T$ is the left Cauchy-Green deformation tensor and the shorthand notation $\Omega_m=\partial\Omega /\partial I_m (m=1,2,4,5,6)$ is adopted here and henceforth. 

In the absence of body forces, free charges and currents, and applying the `quasi-electrostatic approximation', the equations of equilibrium read
\begin{equation}\label{static-governing1}
\div\vec\tau=\vec 0,
\quad
\curl\vec E=\vec 0,
\quad
\div \vec D=0,
\end{equation}
where `$\div$' and `$\curl$' are the divergence and curl operators defined in the deformed configuration, respectively. 

In this paper, we consider an initially stress-free dielectric slab, with flexible electrodes glued to its upper and bottom surfaces, which is bent into a circular sector by the combined action of electric voltage and mechanical loadings. In this case, the electric field in the body is distributed radially in the deformed configuration and there is no exterior electric field in the surrounding vacuum. Then the fields must satisfy the following boundary conditions on the bent surfaces,
\begin{equation}\label{boundary}
\vec\tau\vec n=\vec t_a,
\quad
\vec E\times\vec n=\vec 0,
\quad
\vec D \cdot \vec n=q_e,
\end{equation}
where $\vec t_a$ is the prescribed mechanical traction per unit area of $\partial B_t$, and $q_e$ is the surface charge density on $\partial B_t$.


\subsection{Incremental motions}


We now superimpose an infinitesimal incremental deformation $\vec{\dot x}$ along with an infinitesimal increment in the electric displacement $\vec{\dot D}_l$. Hereinafter, dotted variables represent incremental quantities. The incremental form of the aforementioned equations can be obtained by Taylor expansions. Hence, the linearized incremental forms of the constitutive relations in Eq. \eqref{constitive} read
\begin{equation}\label{incremental-constitive}
\vec{\dot T}_0=\vec A_0\vec H +\vec \Gamma_0\vec{\dot D}_{l0}+p\vec H-\dot{p}\vec I,
\quad
\vec{\dot E}_{l0}=\vec \Gamma_0^T\vec H+\vec K_0\vec{\dot D}_{l0},
\end{equation}
where $\vec{\dot T}_0=\vec F\vec{\dot T}, \vec{\dot E}_{l0}=\vec F^{-\text T}\vec{\dot E}_l$ and $\vec{\dot D}_{l0}=\vec F\vec{\dot D}_l$ are the `push forward' versions of $\vec{\dot T}, \vec{\dot E}_l$ and $\vec{\dot D}_l$, respectively, $\vec H=\grad \vec u$ is the displacement gradient, with $\vec u(\vec x,t)=\vec{\dot x}(\vec X,t)$ being the incremental mechanical displacement, and $\vec A_0, \vec \Gamma_0$ and $\vec K_0$ are, respectively, fourth-, third- and second-order tensors, with Cartesian components defined by
\begin{align}\label{electro-elastic}
&A_{0piqj}=A_{0qjpi}=F_{p\alpha}F_{q\beta}\frac{\partial^2\Omega}{\partial F_{i\alpha}\partial F_{j\beta}},
\quad
\Gamma_{0piq}=\Gamma_{0ipq}=F_{p\alpha}F_{\beta q}^{-1}\frac{\partial^2\Omega}{\partial F_{i\alpha}\partial D_{l\beta}},\notag\\
&K_{0ij}=K_{0ji}=F_{\alpha i}^{-1}F_{\beta j}^{-1}\frac{\partial^2\Omega}{\partial D_{l\alpha}\partial D_{l\beta}}.
\end{align}

The above defined tensors are the so-called `electro-elastic moduli tensors', which are fully determined once the energy function $\Omega$ and biasing fields $\vec F$ and $\vec D_l$ are prescribed.

It is worth noting here that we have the connection
\begin{equation}\label{symmetric}
A_{0jilk}-A_{0ijlk}=\left(\tau_{jl}+p\delta_{jl}\right)\delta_{ik}-\left(\tau_{il}+p\delta_{il}\right)\delta_{jk},
\end{equation}
which can be established by using the incremental form of the symmetry condition of the Cauchy stress $\vec F\vec T=(\vec T\vec F)^{\text T}$.

The incremental forms of the equilibrium equations in \eqref{static-governing1} are
\begin{equation}\label{incremental constitive}
\div\vec{\dot T}_0=\vec 0,
\quad
\curl\vec{\dot E_{l0}}=\vec 0,
\quad
\div \vec{\dot D_{l0}}=0.
\end{equation}

In addition, the incremental incompressibility constraint relation reads
\begin{equation}\label{incompressibility}
\div\vec u=\tr\vec H=0.
\end{equation}

Accordingly, the incremental forms of the boundary conditions \eqref{boundary} are
\begin{equation}\label{incremental-boundary}
\vec{\dot  {T}}_0^{\text T}\vec n=\vec{\dot t}_{A0},
\quad
\vec{\dot E}_{l0}\times\vec n=\vec 0,
\quad
\vec{\dot D}_{l0} \cdot \vec n=\dot q_e,
\end{equation}
where $\vec{\dot t}_{A0}$ and $\dot q_e$ are the incremental mechanical traction and surface charge density per unit area of $\partial B_t$, respectively.


\section{Finite bending and associated stability analysis}
\label{section3}



\subsection{Finite bending deformation}


\begin{figure}[h!]
\centering
\includegraphics[width=0.8\textwidth]{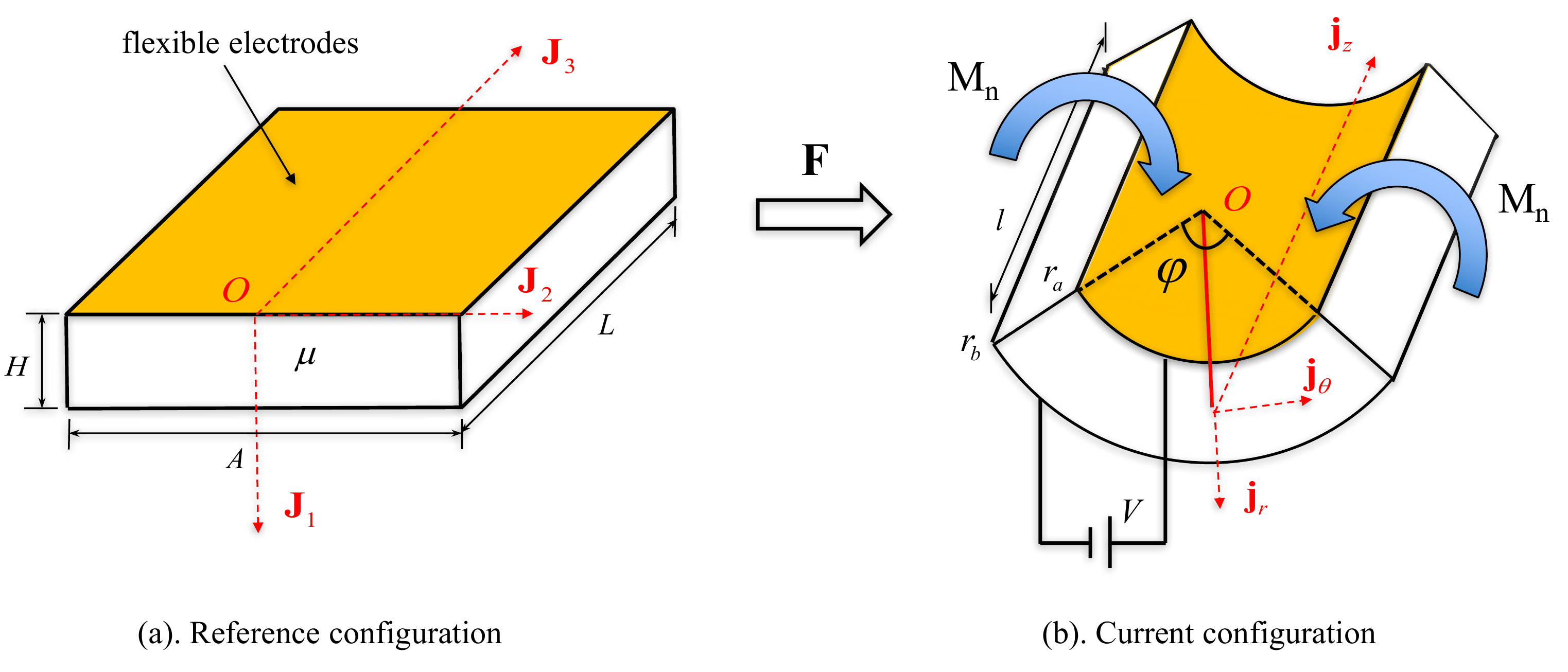}
\caption{
{\footnotesize
Sketch of a dielectric slab with a voltage applied across its thickness subject to finite bending.
}
}
\label{figure2}
\end{figure}

\color{black} 
We consider an initially undeformed dielectric slab of length $L$, thickness $H$ and width $A$, with two flexible electrodes (carbon grease for example) glued onto its top and bottom faces. 
We assume the electrodes to be so thin and soft that their mechanical role can be ignored during the deformation. 
\color{black} 
The width and length aspect ratios of the slab are $A/H$ and $L/H$, respectively. The slab originally occupies the region
\begin{equation}
0 \le X_1 \le H,
\quad
-\frac{A}{2} \le X_2 \le \frac{A}{2},
\quad
0 \le X_3 \le L,
\end{equation}
as depicted in Figure \ref{figure2}(a). With the application of a voltage through the thickness and of mechanical loads (later calculations show that only moments are needed for the bending), the slab bends into the current region
\begin{equation}
r_a \le r \le r_b,
\quad
-\frac{\varphi}{2} \le \theta \le \frac{\varphi}{2},
\quad
0 \le z \le l,
\end{equation}
as depicted in Figure \ref{figure2}(b), through the following bending deformation \citep{Green54, Ogden97}
\begin{equation}\label{bending-deformation}
r=\sqrt{d+\frac{2 X_1}{\omega}},
\quad
\theta=\frac{\omega X_2}{\lambda_z},
\quad
z=\lambda_z X_3,
\end{equation}
where $(X_1,X_2,X_3)$ and $(r,\theta, z)$ are the rectangular Cartesian and cylindrical coordinates in the reference and deformed configurations, with orthogonal bases $(\vec J_1,\vec J_2,\vec J_3)$ and $(\vec j_r,\vec j_\theta,\vec j_z)$, respectively. In Eq. \eqref{bending-deformation}, $d$ and $\omega$ are constants to be determined, $\lambda_z$ is the axial principal stretch, which is taken to be prescribed, $l, r_a, r_b$ and $\varphi$ are the length, inner and outer radii and the bending angle of the deformed sector, respectively, given by
\begin{equation}\label{parameters}
l=\lambda_z L,
\quad
r_a=\sqrt{d},
\quad
r_b=\sqrt{d+\frac{2H}{\omega}},
\quad
\varphi=\frac{\omega A}{\lambda_z}.
\end{equation}

Then the deformation gradient has the following components in the $\vec J_i \bigotimes \vec j_\alpha (i=1,2,3$ and $\alpha=r,\theta,z)$ basis,
\begin{equation}\label{deformation-gradient}
 \vec F = \left[ \begin{matrix}
   \lambda^{-1}\lambda_z^{-1} & 0 & 0  \\[4pt]
   0 & \lambda & 0  \\[4pt]
   0 & 0 & \lambda_z  
\end{matrix} \right],
\end{equation} 
with $\lambda=\omega r/\lambda_z$ being the circumferential principal stretch. Combining Eqs. \eqref{parameters} and \eqref{deformation-gradient}, we establish the following relationships,
\begin{equation}\label{structural-constants}
\omega=\frac{\lambda_z^2\left(\lambda_b^2-\lambda_a^2\right)}{2H}, 
\quad
\varphi=\frac{\lambda_z\left(\lambda_b^2-\lambda_a^2\right)}{2}\frac{A}{H},
\end{equation} 
where $\lambda_a=\omega r_a/\lambda_z,\lambda_b=\omega r_b/\lambda_z$ are the circumferential stretches of the inner and outer surfaces of the deformed sector, respectively.

Now assume that the nominal electric field and electric displacement in the reference configuration are transverse,
\begin{equation}
 \vec E_l = \left[ \begin{matrix}
  E_0 & 0 & 0  
\end{matrix} \right]^{\text T},
\quad
 \vec D_l = \left[ \begin{matrix}
  D_0 & 0 & 0  
\end{matrix} \right]^{\text T},
\end{equation} 
where $E_0$ and $D_0$ are the only non-zero components of the nominal electric field and electric displacement, respectively. Then the true electric field and electric displacement in the deformed configuration are
\begin{equation}
 \vec E=\vec F^{-T}\vec E_l = \left[ \begin{matrix}
  E_r & 0 & 0  
\end{matrix} \right]^{\text T},
\quad
 \vec D=\vec F\vec D_l = \left[ \begin{matrix}
  D_r & 0 & 0  
\end{matrix} \right]^{\text T},
\end{equation} 
where  $E_r=\lambda\lambda_zE_0=E_0\omega r$ and $D_r=\lambda^{-1}\lambda_z^{-1}D_0=D_0/(\omega r)$. The Maxwell equation \eqref{static-governing1}$_3$ reads 
\begin{equation}
\frac{\partial D_r}{\partial r}+\frac{1}{r}D_r=\frac{1}{r}\frac{\partial (rD_r)}{\partial r}=0,
\end{equation} 
showing that $D_0$ is a constant. Notice, however, that $E_0$ is not a constant.

According to Equation \eqref{invariants}, the invariants are
\begin{align}
&I_1=\lambda^2+\lambda^{-2}\lambda_z^{-2}+\lambda_z^2,
\quad
I_2=\lambda^{-2}+\lambda^{2}\lambda_z^{2}+\lambda_z^{-2},
\quad\notag\\
&I_4=D_0^2,
\quad
I_5=\lambda^{-2}\lambda_z^{-2}D_0^2,
\quad
I_6=\lambda^{-4}\lambda_z^{-4}D_0^2.
\end{align}
From Eqs. \eqref{stress} and \eqref{electric}, we further obtain the non-zero components of the Cauchy stress $\vec \tau$ and of the electric field $\vec E$ as
\begin{align}\label{stress-component}
&\tau_{rr}=2\lambda^{-2}\lambda_z^{-2}\Omega_1+2\left(\lambda^{-2}+\lambda_z^{-2}\right)\Omega_2+2\lambda^{-2}\lambda_z^{-2}D_0^2\Omega_5+4\lambda^{-4}\lambda_z^{-4}D_0^2\Omega_6-p,
\quad\notag\\
&\tau_{\theta\theta}=2\lambda^{2}\Omega_1+2\left(\lambda_z^{-2}+\lambda^2\lambda_z^{2}\right)\Omega_2-p,
\quad\notag\\
&\tau_{zz}=2\lambda_z^{2}\Omega_1+2\left(\lambda^{-2}+\lambda^2\lambda_z^{2}\right)\Omega_2-p,
\end{align}
\begin{align}\label{electric-component}
E_r & = 2\left(\lambda^2\lambda_z^2\Omega_4+\Omega_5+\lambda^{-2}\lambda_z^{-2}\Omega_6\right)D_r \notag \\
&  =2\left(\lambda\lambda_z\Omega_4+\lambda^{-1}\lambda_z^{-1}\Omega_5+\lambda^{-3}\lambda_z^{-3}\Omega_6\right)D_0 .
\end{align}

At this stage we note that the energy function has only three independent variables: $\lambda,\lambda_z$ and $D_0$. Introducing a reduced energy function $W$ defined by
\begin{equation}
W\left(\lambda,\lambda_z,D_0\right)=\Omega\left(I_1,I_2,I_4,I_5,I_6\right).
\end{equation}

Eqs. \eqref{stress-component} and \eqref{electric-component} can be rewritten compactly as
\begin{equation}\label{simple-stress}
\tau_{rr}-\tau_{\theta\theta}=-\lambda\frac{\partial W}{\partial\lambda},
\end{equation}
\begin{equation}\label{electric-components}
E_0=\lambda^{-1}\lambda_z^{-1}E_r=\frac{\partial W}{\partial D_0}.
\end{equation}

For the considered deformation, the equilibrium equation \eqref{static-governing1}$_1$ reduces to the radial component equation
\begin{equation}\label{balance}
\frac{\partial \tau_{rr}}{\partial r}+\frac{1}{r}\left(\tau_{rr}-\tau_{\theta\theta}\right)=0.
\end{equation}

Combining Eqs. \eqref{simple-stress} and \eqref{balance} and using the relation $\text{d}\lambda/\lambda=\text{d}r/r$ enables us to rewrite the principal stress components $\tau_{rr}$ and $\tau_{\theta\theta}$ as
\begin{equation}\label{normal-stress}
\tau_{rr}=W+K,
\end{equation}
\begin{equation}\label{shear-stress}
\tau_{\theta\theta}=\lambda\frac{\partial W}{\partial \lambda}+\tau_{rr}=\lambda\frac{\partial W}{\partial \lambda}+W+K,
\end{equation}
where $K$ is a constant to be determined from the boundary conditions. Here the inner and outer surfaces at $r_a$ and $r_b$ are free of mechanical tractions, so that
\begin{equation}\label{static-boundary}
\tau_{rr}(r_a)=\tau_{rr}(r_b)=0.
\end{equation}
Then the constant $K$ can be obtained as
\begin{equation}\label{constant-K}
K=-W\left(\lambda_a,\lambda_z,D_0\right)=-W\left(\lambda_b,\lambda_z,D_0\right),
\end{equation}
and the connection between $\lambda_a,\lambda_b,\lambda_z$ and $D_0$ can be established as
\begin{equation}\label{ab}
W\left(\lambda_a,\lambda_z,D_0\right)-W\left(\lambda_b,\lambda_z,D_0\right)=0.
\end{equation}

According to Eq. \eqref{static-governing1}$_{2}$, the electric field can be expressed as $\vec E=-\grad\phi$, where $\phi$ is the electric potential, with the only non-zero radial electric field component given by $E_r=-d\phi/dr$. We denote the electric voltage difference between the inner and outer surfaces as $V=\phi_a-\phi_b$, which, with the help of Eqs. \eqref{structural-constants}$_1$ and \eqref{electric-components}, can be obtained as
\begin{equation}\label{connection}
V=\int_{r_a}^{r_b}\lambda\lambda_z\frac{\partial W}{\partial D_0}\,\text dr
=\frac{\lambda_z^2}{\omega}\int_{\lambda_a}^{\lambda_b}\lambda\frac{\partial W}{\partial D_0}\,\text d\lambda
=\frac{2H}{\lambda_b^2-\lambda_a^2}\int_{\lambda_a}^{\lambda_b}\lambda\frac{\partial W}{\partial D_0}\,\text d\lambda.
\end{equation}
Eq. \eqref{connection} provides the equilibrium relation between the constants $V,D_0,\lambda_a,\lambda_b$ and $\lambda_z$, once the energy function of the material  is specified.

Then by solving the Eqs. \eqref{structural-constants}$_2$, \eqref{ab} and \eqref{connection}, $\lambda_a,\lambda_b$ and $D_0$ can be determined once $V,\varphi,\lambda_z$ and $A/H$ are given. Eventually the inner and outer radii of the deformed sector $r_a$ and $r_b$, the constant $\omega$ and the circumferential principal stretch of arbitrary point in the sector $\lambda$ can be derived as
\begin{equation}
r_a=\frac{\lambda_a\lambda_z}{\omega}=\frac{\lambda_a}{\varphi}A,
\quad
r_b=\frac{\lambda_b\lambda_z}{\omega}=\frac{\lambda_b}{\varphi}A,
\quad
\omega=\frac{\lambda_z\varphi}{A},
\quad
\lambda=\frac{\omega r}{\lambda_z}.
\end{equation}

As a result, the configuration and the distributions of stretches of the deformed sector are fully determined. Finally, the required applied axial force $F_N$ and the moment $M_n$ about the origin on the lateral faces $\theta=\pm\omega A/(2\lambda_z)$ can be determined as
\begin{align}\label{force1}
&F_N=\lambda_zL\int_{r_a}^{r_b}\tau_{\theta\theta}\,\text dr
=\frac{2HL}{\lambda_b^2-\lambda_a^2}\int_{\lambda_a}^{\lambda_b}\tau_{\theta\theta}\,\text d\lambda
=\mu HL\overline F_N,\notag\\
&M_n=\lambda_zL\int_{r_a}^{r_b}r\tau_{\theta\theta}\,\text dr
=\frac{4H^2L}{\lambda_z\left(\lambda_b^2-\lambda_a^2\right)^2}\int_{\lambda_a}^{\lambda_b}\lambda\tau_{\theta\theta}\,\text d\lambda
=\mu H^2L\overline M_n,
\end{align}
where $\mu$ is the initial mechanical shear modulus, $\overline F_N$ and $\overline M_n$ are dimensionless measures of the axial force and moment, respectively. Note that from Eq. \eqref{balance} we have the relation $\tau_{\theta\theta}=\text d (r\tau_{rr})/(\text d r)$, thus Eq. \eqref{force1}$_1$ reads
\begin{equation}\label{normal-force}
F_N=\frac{2HL}{\lambda_b^2-\lambda_a^2}\int_{r_a}^{r_b}\tau_{\theta\theta}\,\text d r
=\frac{2HL}{\lambda_b^2-\lambda_a^2}\left[r_b\tau_{rr}(r_b)-r_a\tau_{rr}(r_a)\right],
\end{equation}
which identically equals to zero due to the boundary condition \eqref{static-boundary}. Hence, only moments are required to bend the slab.


\subsection{Small-amplitude wrinkle}


We now superimpose a small harmonic inhomogeneous deformation on the underlying deformed configuration of the sector, to model the onset of wrinkling on the inner curved face. 

We start with the components of the incremental displacement and the incremental electric displacement in the form 
\begin{equation}
u_i=u_i(r,\theta,z),\quad \dot{D}_{l0i}=\dot{D}_{l0i}(r,\theta,z).
\end{equation}
Then the incremental displacement gradient reads
\begin{equation}
 \vec H = \left[ \begin{matrix}
   \frac{\partial u_r}{\partial r} & \frac{1}{r}\left(\frac{\partial u_r}{\partial \theta}-u_\theta\right) & \frac{\partial u_r}{\partial z}  \\[4pt]
   \frac{\partial u_\theta}{\partial r} & \frac{1}{r}\left(\frac{\partial u_\theta}{\partial \theta}+u_r\right) & \frac{\partial u_\theta}{\partial z}  \\[4pt]
   \frac{\partial u_z}{\partial r} & \frac{1}{r}\frac{\partial u_z}{\partial \theta} & \frac{\partial u_z}{\partial z}  
\end{matrix} \right],
\end{equation} 
in the $\vec j_\alpha \otimes \vec j_\beta (\alpha,\beta=r,\theta,z)$ basis, and the incompressibility condition Eq. \eqref{incompressibility} for the incremental motion reads
\begin{equation}\label{incremental-incompressibility}
\div\vec u=\tr \vec H=\frac{\partial u_r}{\partial r}+\frac{1}{r}\left(\frac{\partial u_\theta}{\partial \theta}+u_r\right)+\frac{\partial u_z}{\partial z}=0.
\end{equation}
From Eq. \eqref{incremental constitive}$_2$, we introduce an incremental electric potential $\dot \phi$, and the components of the incremental electric field are
\begin{equation}
\dot{E}_{l0r}=-\frac{\partial \dot \phi}{\partial r},
\quad
\dot{E}_{l0\theta}=-\frac{1}{r}\frac{\partial \dot \phi}{\partial \theta},
\quad
\dot{E}_{l0z}=-\frac{\partial \dot \phi}{\partial z}.
\end{equation}

Now the electro-elastic moduli tensors $\vec A_0,\vec {\Gamma}_0$ and $\vec K_0$ can be evaluated according to Eq. \eqref{electro-elastic}, with non-zero components listed in Appendix A. Then the components of the incremental stress and electric fields are expanded as \citep{Wu17}
\begin{align}\label{incremental-constitutive1}
&\dot T_{0rr}=\left(A_{01111}+p\right)\frac{\partial u_r}{\partial r}+A_{01122}\frac{1}{r}\left(\frac{\partial u_\theta}{\partial \theta}+u_r\right)+A_{01133}\frac{\partial u_z}{\partial z}+\Gamma_{0111}\dot D_{l0r}-\dot p,\notag \\
&\dot T_{0\theta\theta}=A_{01122}\frac{\partial u_r}{\partial r}+\left(A_{02222}+p\right)\frac{1}{r}\left(\frac{\partial u_\theta}{\partial \theta}+u_r\right)+A_{02233}\frac{\partial u_z}{\partial z}+\Gamma_{0221}\dot D_{l0r}-\dot p,\notag \\
&\dot T_{0zz}=A_{01133}\frac{\partial u_r}{\partial r}+A_{02233}\frac{1}{r}\left(\frac{\partial u_\theta}{\partial \theta}+u_r\right)+\left(A_{03333}+p\right)\frac{\partial u_z}{\partial z}+\Gamma_{0331}\dot D_{l0r}-\dot p,\notag \\
&\dot T_{0r\theta}=A_{01212}\frac{\partial u_\theta}{\partial r}+\left(A_{01221}+p\right)\frac{1}{r}\left(\frac{\partial u_r}{\partial \theta}-u_\theta\right)+\Gamma_{0122}\dot D_{l0\theta},\notag \\
&\dot T_{0rz}=A_{01313}\frac{\partial u_z}{\partial r}+\left(A_{01331}+p\right)\frac{\partial u_r}{\partial z}+\Gamma_{0133}\dot D_{l0z},\notag \\
&\dot T_{0\theta r}=A_{02121}\frac{1}{r}\left(\frac{\partial u_r}{\partial \theta}-u_\theta\right)+\left(A_{01221}+p\right)\frac{\partial u_\theta}{\partial r}+\Gamma_{0122}\dot D_{l0\theta},\notag \\
&\dot T_{0\theta z}=A_{2323}\frac{1}{r}\frac{\partial u_z}{\partial \theta}+(A_{02332}+p)\frac{\partial u_\theta}{\partial z},\notag \\
&\dot T_{0zr}=A_{03131}\frac{\partial u_r}{\partial z}+(A_{01331}+p)\frac{\partial u_z}{\partial r}+\Gamma_{0133}\dot D_{l0z},\notag \\
&\dot T_{0z\theta}=A_{03232}\frac{\partial u_\theta}{\partial z}+(A_{02332}+p)\frac{1}{r}\frac{\partial u_z}{\partial \theta},
\end{align}
and 
\begin{align}\label{incremental-constitutive2}
&\dot E_{l0r}=-\frac{\partial\dot{\phi}}{\partial r}=\Gamma_{0111}\frac{\partial u_r}{\partial r}+\Gamma_{221}\frac{1}{r}\left(\frac{\partial u_\theta}{\partial \theta}+u_r\right)+\Gamma_{0331}\frac{\partial u_z}{\partial z}+K_{011}\dot D_{l0r},\notag\\
&\dot E_{l0\theta}=-\frac{1}{r}\frac{\partial\dot{\phi}}{\partial \theta}=\Gamma_{0122}\left[\frac{1}{r}\left(\frac{\partial u_r}{\partial \theta}-u_\theta\right)+\frac{\partial u_\theta}{\partial r}\right]+K_{022}\dot D_{l0\theta},\notag\\
&\dot E_{l0z}=-\frac{\partial\dot{\phi}}{\partial z}=\Gamma_{0133}\left(\frac{\partial u_r}{\partial z}+\frac{\partial u_z}{\partial_r}\right)+K_{033}\dot D_{l0z},
\end{align}
according to Eq. \eqref{incremental-constitive}.

Finally, the incremental forms of equilibrium equation \eqref{incremental constitive}$_1$ and the incremental Maxwell equation \eqref{incremental constitive}$_3$ reduce to
\begin{align}\label{incremental-constitutive3}
&\frac{\partial \dot T_{0rr}}{\partial r}+\frac{1}{r}\frac{\partial \dot T_{0\theta r}}{\partial \theta}+\frac{\dot T_{0rr}-\dot T_{0\theta\theta}}{r}+\frac{\partial \dot T_{0zr}}{\partial z}=0,\notag\\
&\frac{\partial \dot T_{0r\theta}}{\partial r}+\frac{1}{r}\frac{\partial \dot T_{0\theta \theta}}{\partial \theta}+\frac{\dot T_{0\theta r}+\dot T_{0r\theta}}{r}+\frac{\partial \dot T_{0z\theta}}{\partial z}=0,\notag\\
&\frac{\partial \dot T_{0rz}}{\partial r}+\frac{1}{r}\frac{\partial \dot T_{0\theta z}}{\partial \theta}+\frac{\partial \dot T_{0zz}}{\partial z}+\frac{\dot T_{0rz}}{r}=0,
\end{align}
and
\begin{equation}\label{incremental-Maxwell}
\frac{\partial \dot D_{l0r}}{\partial r}+\frac{1}{r}\left(\frac{\partial \dot D_{l0\theta}}{\partial \theta}+\dot D_{l0r}\right)+\frac{\partial \dot D_{l0z}}{\partial z}=0,
\end{equation}
respectively.

We assume that the sector is under end thrust at the lateral faces $\theta=\pm\omega A/(2\lambda_z)$ and $z=0,l$, while the two surfaces $r=r_a,r_b$ remain traction-free and the applied voltage is taken to be a constant. The boundary conditions for the incremental fields are
\begin{align}\label{boundary1}
& u_\theta=\dot T_{0\theta r}=\dot T_{0\theta z}=0 \quad & \text{at} & \quad \theta=\pm\omega A/(2\lambda_z), \notag \\
& u_z=\dot T_{0zr}=\dot T_{0z\theta}=0 \quad & \text{at} & \quad z=0,\lambda_zL,\notag \\
& \dot T_{0rr}=\dot T_{0r\theta}=\dot T_{0rz}=\dot \phi=0 \quad & \text{at} & \quad r=r_a,r_b.
\end{align}


\subsection{Stroh formulation}


We seek solutions of equations in Section 3.2 in the form \citep{Su16b}
\begin{align}\label{incremental-solutions}
&u_r=U_r(r)\text{cos}\left(n\theta\right)\text{cos}\left(kz\right), & \quad & u_\theta=U_\theta(r)\text{sin}\left(n\theta\right)\text{cos}\left(kz\right), \notag \\
&u_z=U_z(r)\text{cos}\left(n\theta\right)\text{sin}\left(kz\right), & \quad & \dot{\phi}=\Phi(r)\text{cos}\left(n\theta\right)\text{cos}\left(kz\right), \notag \\
&\dot T_{0rr}=\Sigma_{rr}(r)\text{cos}\left(n\theta\right)\text{cos}\left(kz\right), & \quad & \dot T_{0r\theta}=\Sigma_{r\theta}(r)\text{sin}\left(n\theta\right)\text{cos}\left(kz\right), \notag \\
&\dot T_{0rz}=\Sigma_{rz}(r)\text{cos}\left(n\theta\right)\text{sin}\left(kz\right), & \quad & \dot D_{l0r}=\Delta_r(r)\text{cos}\left(n\theta\right)\text{cos}\left(kz\right),
\end{align}
where $n$ and $k$ are the circumferential and axial wave numbers, respectively. Then from the incremental constitutive equations \eqref{incremental-constitutive1}, \eqref{incremental-constitutive2} and the incremental boundary conditions \eqref{boundary1}$_{1,2}$, we have
\begin{equation}
n=\frac{2\lambda_zq\pi}{\omega A}=\frac{4q\pi}{\lambda_z\left(\lambda_b^2-\lambda_a^2\right)}\frac{H}{A},\quad
k=\frac{m\pi}{\lambda_z L} \quad (q,m=0,1,2,...),
\end{equation}
where the positive integers $q$ and $m$ give the numbers of circumferential and axial wrinkles of the sector, respectively \citep{Destrade09,Balbi15}. It should be noticed that they cannot be zero simultaneously.

Then Eqs. \eqref{incremental-incompressibility}-\eqref{incremental-Maxwell} that govern the incremental motion of the dielectric sector can be rearranged to yield the following first-order differential system \citep{Gilchrist09, Destrade09, Destrade14, Balbi15}
\begin{equation}\label{Stroh}
\frac{\text{d}}{\text{d}r}\vec \eta(r)=\frac{1}{r}\vec G(r)\vec\eta(r),
\end{equation}
where
\begin{equation}\label{Stroh-vector}
\vec\eta(r)=\left[ \begin{matrix}
 U_r & U_\theta & U_z & r\Delta_r & r\Sigma_{rr} &r\Sigma_{r\theta} & r\Sigma_{rz} & \Phi 
\end{matrix} \right]^{\text T}=\left[ \begin{matrix}\vec U & \vec S\end{matrix} \right]^{\text T},
\end{equation}
is the Stroh vector (with $\vec U=\left[ \begin{matrix}
 U_r & U_\theta & U_z & r\Delta_r\end{matrix} \right]^{\text T}$ and $\vec S=\left[ \begin{matrix}
r\Sigma_{rr} &r\Sigma_{r\theta} & r\Sigma_{rz} & \Phi 
\end{matrix} \right]^{\text T}$), $\vec G$ is the so-called Stroh matrix, which has the following block structure
\begin{equation}
\vec G=\left[ \begin{matrix}
 \vec G_1 & \vec G_2 \\
 \vec G_3 & \vec G_4\end{matrix} \right],
\end{equation}
where the four $4\times 4$ sub-blocks $\vec G_1, \vec G_2, \vec G_3$ and $\vec G_4$ have the following components
\begin{align}
&\vec G_1=\left[ \begin{matrix}
 -1 & -n & -kr & 0\\
 \frac{n\left(\gamma_{12}-\tau_{rr}\right)}{\gamma_{12}} &  \frac{\gamma_{12}-\tau_{rr}}{\gamma_{12}} & 0 & 0\\
 \frac{kr\left(\gamma_{13}-\tau_{rr}\right)}{\gamma_{13}} & 0 & 0 & 0 \\
 \xi_1 & -\frac{n\tau_{rr}}{\gamma_{12}}\frac{\Gamma_{0122}}{K_{022}} & 0 & 0
\end{matrix} \right], \quad
\vec G_2=\left[\begin{matrix}
0 & 0 & 0 & 0 \\
0 & \frac{1}{\gamma_{12}} & 0 & -\frac{n}{\gamma_{12}}\frac{\Gamma_{0122}}{K_{022}} \\
0 & 0 & \frac{1}{\gamma_{13}} & -\frac{kr}{\gamma_{13}}\frac{\Gamma_{0133}}{K_{033}}\\
0 & \frac{n}{\gamma_{12}}\frac{\Gamma_{0122}}{K_{022}} & \frac{kr}{\gamma_{13}}\frac{\Gamma_{0133}}{K_{033}} & \xi_2
\end{matrix}\right], \notag \\
&\vec G_3=\left[\begin{matrix}
\kappa_{11} & \kappa_{12} & \kappa_{13} &-\left(\Gamma_{0111}-\Gamma_{0221}\right) \\
\kappa_{12} & \kappa_{22} & \kappa_{23} &-n\left(\Gamma_{0111}-\Gamma_{0221}\right) \\
\kappa_{13} & \kappa_{23} & \kappa_{33} &-kr\left(\Gamma_{0111}-\Gamma_{0331}\right)\\
\Gamma_{0111}-\Gamma_{0221} & n\left(\Gamma_{0111}-\Gamma_{0221}\right) & kr\left(\Gamma_{0111}-\Gamma_{0331}\right) & -K_{011}
\end{matrix}\right], \notag \\
&\vec G_4=\left[ \begin{matrix}
 1 & -\frac{n\left(\gamma_{12}-\tau_{rr}\right)}{\gamma_{12}} & -\frac{kr\left(\gamma_{13}-\tau_{rr}\right)}{\gamma_{13}} & \xi_1\\
 n &  -\frac{\gamma_{12}-\tau_{rr}}{\gamma_{12}} & 0 & -\frac{n\tau_{rr}}{\gamma_{12}}\frac{\Gamma_{0122}}{K_{022}}\\
 kr & 0 & 0 & 0 \\
 0 & 0 & 0 & 0
\end{matrix} \right].
\end{align}
Here
\begin{align}\label{material-parameters}
&\gamma_{12}=A_{01212}-\frac{\Gamma_{0122}^2}{K_{022}}, \quad \gamma_{21}=A_{02121}-\frac{\Gamma_{0122}^2}{K_{022}}, \quad \gamma_{23}=A_{02323}, \notag \\
&\gamma_{13}=A_{01313}-\frac{\Gamma_{0133}^2}{K_{033}}, \quad \gamma_{31}=A_{03131}-\frac{\Gamma_{0133}^2}{K_{033}}, \quad \gamma_{32}=A_{03232},\notag \\
&\xi_1=-\left(\frac{\Gamma_{0122}}{K_{022}}\frac{n^2}{\gamma_{12}}+\frac{\Gamma_{0133}}{K_{033}}\frac{k^2r^2}{\gamma_{13}}\right)\tau_{rr}, \notag \\
&\xi_2=-\left(\frac{n^2}{K_{022}}+\frac{\Gamma_{0122}^2}{K^2_{022}}\frac{n^2}{\gamma_{12}}+\frac{k^2r^2}{K_{033}}+\frac{\Gamma_{0133}^2}{K_{033}^2}\frac{k^2r^2}{\gamma_{13}}\right), \notag \\
& \beta_{12}=\frac{1}{2}\left(A_{01111}+A_{02222}-2A_{01122}-2A_{01221}+\frac{2\Gamma_{0122}^2}{K_{022}}\right), \notag \\
&\beta_{13}=\frac{1}{2}\left(A_{01111}+A_{03333}-2A_{01133}-2A_{01331}+\frac{2\Gamma_{0133}^2}{K_{033}}\right), \notag \\
&\kappa_{11}=2(\gamma_{12}-\tau_{rr}+\beta_{12})+n^2\left[\gamma_{21}-\frac{\left(\gamma_{12}-\tau_{rr}\right)^2}{\gamma_{12}}\right]+k^2r^2\left[\gamma_{31}-\frac{\left(\gamma_{13}-\tau_{rr}\right)^2}{\gamma_{13}}\right], \notag \\
&\kappa_{12}=n\left(\gamma_{12}+\gamma_{21}+2\beta_{12}-\frac{\tau_{rr}^2}{\gamma_{12}}\right), \quad
\kappa_{13}=kr\left(A_{01111}+A_{02233}-A_{01122}-A_{01133}+p\right),\notag \\
&\kappa_{22}=2n^2(\gamma_{12}-\tau_{rr}+\beta_{12})+\gamma_{21}-\frac{\left(\gamma_{12}-\tau_{rr}\right)^2}{\gamma_{12}}+k^2r^2\gamma_{32}, \notag \\
&\kappa_{23}=nkr\left(A_{01111}+A_{02233}+A_{02332}-A_{01122}-A_{01133}+2p\right),\notag \\
&\kappa_{33}=2k^2r^2\left(\gamma_{13}-\tau_{rr}+\beta_{13}\right)+n^2\gamma_{23}.
\end{align}

It should be noticed that to derive Eqs. \eqref{Stroh}-\eqref{material-parameters}, we made use of the connections
\begin{equation}
A_{01221}+p=A_{01212}-\tau_{rr}, \quad A_{01331}+p=A_{01313}-\tau_{rr},
\end{equation}
which result from Eqs. \eqref{electro-elastic}$_1$ and \eqref{symmetric}. The derivation of the Stroh formulation is given in Appendix B.

Now the incremental boundary conditions \eqref{boundary1}$_3$ read
\begin{equation}\label{boundary-Stroh}
\vec S(r_a)=\vec S(r_b)=\vec 0.
\end{equation}

Note that we chose to write the components of $\vec \eta$ in the order presented in Eq. \eqref{Stroh-vector}, because it will turn out to be the most practical for those boundary value problems where the electric field is due to a constant voltage applied to the bent faces of the sector. For the case where the sector is charge-controlled \citep{Keplinger10, Dorf14a, Su16a, Su16b} instead of voltage-controlled, the places of $r\Delta$ and $\Phi$ must be swapped in $\vec \eta$ for greater efficiency in the scheme. In other words, $\vec \eta$ and $\vec G$ in Eq. \eqref{Stroh} should be replaced with $\hat{\vec \eta}$ and $\hat{\vec G}$, respectively, where
 \begin{equation}
 \hat{\vec \eta} = \left[\begin{matrix}
 U_r \\ U_{\theta} \\U_z \\ \Phi \\ r\Sigma_{rr} \\ r\Sigma_{\theta\theta} \\ r\Sigma_{zz}\\ r\Delta
 \end{matrix}\right]=\left[\begin{matrix}\hat{\vec U} \\ \hat{\vec S}\end{matrix}\right]=\vec{R}\vec{\eta},
\qquad
 \hat{\vec G} = \vec{RGR}^{-1}, \qquad
\vec{R}=\left[\begin{matrix}
1 & 0 & 0 & 0 & 0 & 0 & 0 & 0 \\
0 & 1 & 0 & 0 & 0 & 0 & 0 & 0 \\
0 & 0 & 1 & 0 & 0 & 0 & 0 & 0 \\
0 & 0 & 0 & 0 & 0 & 0 & 0 & 1 \\
0 & 0 & 0 & 0 & 1 & 0 & 0 & 0 \\
0 & 0 & 0 & 0 & 0 & 1 & 0 & 0 \\
0 & 0 & 0 & 0 & 0 & 0 & 1 & 0 \\
0 & 0 & 0 & 1 & 0 & 0 & 0 & 0 \\
\end{matrix}\right],
\end{equation}
with $\hat{\vec U}=\left[ \begin{matrix}
 U_r & U_\theta & U_z & \Phi\end{matrix} \right]^{\text T}$ and $\hat{\vec S}=\left[ \begin{matrix}
r\Sigma_{rr} &r\Sigma_{r\theta} & r\Sigma_{rz} & r\Delta_r 
\end{matrix} \right]^{\text T}$. Then the traction-free boundary conditions at the two surfaces $r_a,r_b$, Eq. \eqref{boundary-Stroh} should be modified as
\begin{equation}
\hat{\vec S}(r_a)=\hat{\vec S}(r_b)=\vec 0.
\end{equation}

As a result, the method presented in this paper can be easily extended to the case of a charge-controlled sector. Our calculations (not presented here) show that we then recover the same results as in the literature when the slab is reduced to a half-space \citep{Dorf10b}.


\subsection{The surface impedance matrix method}


The inhomogeneous differential system \eqref{Stroh} is stiff numerically, especially for thick slabs. Over the years, several algorithms such as the compound matrix method \citep{Shmuel13} and the state space method \citep{Wu17} have been adopted to overcome the stiffness of this equation. Here the so-called surface impedance matrix method \citep{Gilchrist09, Destrade09, Destrade14, Balbi15} is employed to build a robust and efficient numerical procedure for obtaining the dispersion equation. 

We introduce the $8\times 8$ matricant $\vec M(r,r_a)=\left[\begin{matrix}\vec M_1(r,r_a) & \vec M_2(r,r_a) \\ \vec M_3(r,r_a)& \vec M_4(r,r_a)\end{matrix}\right]$, which is defined as the matrix such that
\begin{equation}
\vec\eta(r)=\vec M(r,r_a)\vec\eta(r_a),
\end{equation}
with the obvious condition that
\begin{equation}\label{matrix-M}
\vec M(r_a,r_a)=\vec I_{8 \times 8}.
\end{equation}

Use of the incremental boundary condition $\vec S({r_a})=0$ gives
\begin{equation}\label{stress-Stroh}
\vec S(r)=\vec z^a(r,r_a)\vec U(r),
\end{equation}
where $\vec z^a(r,r_a)$ is the conditional impedance matrix, which is defined as
\begin{equation}\label{matrix-z}
\vec z^a(r,r_a)=\vec M_3(r,r_a)\vec M_1^{-1}(r,r_a).
\end{equation}

Substituting Eq. \eqref{stress-Stroh} into Eq. \eqref{Stroh} gives
\begin{equation}\label{two_Riccati}
\frac{\text d}{\text dr}\vec U=\frac{1}{r}\vec G_1 \vec U+\frac{1}{r}\vec G_2 \vec z^a \vec U, \quad
\frac{\text d}{\text dr}(\vec z^a \vec U)=\frac{1}{r}\vec G_3 \vec U+\frac{1}{r}\vec G_4 \vec z^a \vec U.
\end{equation}

Elimination of $\vec U$ from Eq. \eqref{two_Riccati} yields the following Riccati differential equation
\begin{equation}\label{Riccati}
\frac{\text d\vec z^a}{\text dr}=\frac{1}{r}\left(-\vec z^a \vec G_1-\vec z^a \vec G_2 \vec z^a+\vec G_3+\vec G_4 \vec z^a\right),
\end{equation}
with the initial condition
\begin{equation}\label{initial-condition1}
\vec z^a(r_a,r_a)=\vec 0,
\end{equation}
which follows from Eqs. \eqref{matrix-M} and \eqref{matrix-z}.

Then we integrate Eq. \eqref{Riccati} numerically with the initial condition \eqref{initial-condition1} from $r_a$ to $r_b$ and tune the bending angle until the following target condition is satisfied
\begin{equation}
\text{det}\ \vec z^a(r_b,r_a)=0,
\end{equation}
which results from the boundary
\begin{equation}\label{boundary-displacement}
\vec S(r_b)=\vec z^a(r_b,r_a)\vec U(r_b)=\vec 0.
\end{equation}

The conclusion is that, for a given voltage $V$, the critical bending angle $\varphi_c$ can be determined, and so can the critical value of the inner circumferential stretch $\lambda_a$, which we denote by $\lambda_c$.

It follows from Eq. \eqref{boundary-displacement} that the ratios of the incremental motion on the outer face of the sector can be determined as
\begin{align}
t_\theta &= \frac{U_\theta(r_b)}{U_r(r_b)} \notag \\
& =\frac{Q_{11}Q_{24}Q_{33}+Q_{13}Q_{21}Q_{34}+Q_{14}Q_{23}Q_{31}-Q_{11}Q_{23}Q_{34}-Q_{13}Q_{24}Q_{31}-Q_{14}Q_{21}Q_{33}}{Q_{12}Q_{23}Q_{34}+Q_{13}Q_{24}Q_{32}+Q_{14}Q_{22}Q_{33}-Q_{12}Q_{24}Q_{33}-Q_{13}Q_{22}Q_{34}-Q_{14}Q_{23}Q_{32}}, \notag \\
t_z & =\frac{U_z(r_b)}{U_r(r_b)} \notag \\
& =\frac{Q_{11}Q_{22}Q_{34}+Q_{12}Q_{24}Q_{31}+Q_{14}Q_{21}Q_{32}-Q_{11}Q_{24}Q_{32}-Q_{12}Q_{21}Q_{34}-Q_{14}Q_{22}Q_{31}}{Q_{12}Q_{23}Q_{34}+Q_{13}Q_{24}Q_{32}+Q_{14}Q_{22}Q_{33}-Q_{12}Q_{24}Q_{33}-Q_{13}Q_{22}Q_{34}-Q_{14}Q_{23}Q_{32}}, \notag \\
t_\Phi &=\frac{\Phi(r_b)}{U_r(r_b)} \notag \\
&=\frac{Q_{11}Q_{23}Q_{32}+Q_{12}Q_{21}Q_{33}+Q_{13}Q_{22}Q_{31}-Q_{11}Q_{22}Q_{33}-Q_{12}Q_{23}Q_{31}-Q_{13}Q_{21}Q_{32}}{Q_{12}Q_{23}Q_{34}+Q_{13}Q_{24}Q_{32}+Q_{14}Q_{22}Q_{33}-Q_{12}Q_{23}Q_{31}-Q_{12}Q_{23}Q_{31}-Q_{13}Q_{21}Q_{32}},
\end{align}
where the shorthand notation $Q_{ij}=z_{ij}^a(r_b,r_a) (i,j=1,2,3,4)$ is used.

On the other hand, we can also start at the outer surface $r=r_b$ and introduce the $8 \times 8$ matricant $\vec M(r,r_b)=\left[\begin{matrix}\vec M_1(r,r_b) & \vec M_2(r,r_b) \\ \vec M_3(r,r_b)& \vec M_4(r,r_b)\end{matrix}\right]$ such that
\begin{equation}
\vec\eta(r)=\vec M(r,r_b)\vec\eta(r_b),
\end{equation}
with the obvious condition that
\begin{equation}
\vec M(r_b,r_b)=\vec I_{8 \times 8}.
\end{equation}

Following the same procedure, we can also obtain a Riccati differential equation for the other conditional impedance matrix $\vec z^b(r,r_b)$, as
\begin{equation}\label{Raccati-b}
\frac{\text d\vec z^b}{\text dr}=\frac{1}{r}\left(-\vec z^b \vec G_1-\vec z^b \vec G_2 \vec z^b+\vec G_3+\vec G_4 \vec z^b\right).
\end{equation}

The corresponding form of Eq. \eqref{two_Riccati}$_1$ is
\begin{equation}\label{two_Riccati-b}
\frac{\text d}{\text dr}\vec U=\frac{1}{r}\vec G_1 \vec U+\frac{1}{r}\vec G_2 \vec z^b\vec U.
\end{equation}

With the critical stretch $\lambda_c$ obtained by integrating the Riccati differential equation for the $\vec z^a(r,r_a)$ conditional impedance matrix, we can now integrate simultaneously Eqs. \eqref{Raccati-b} and \eqref{two_Riccati-b} from $r_b$ to $r_a$ with the following initial conditions
\begin{equation}
\vec U(r_b)=U(r_b)\left[\begin{matrix}1 & t_\theta & t_z & t_\Phi \end{matrix}\right]^{\text T}, \quad \vec z^b(r_b, r_b)=\vec 0,
\end{equation}
to determine the full distribution of the incremental field $\vec U$ in the deformed sector and corresponding buckling pattern.


 \section{Numerical results and discussion}
 \label{section4}


For illustration, we now consider the so-called ideal neo-Hookean dielectric model:
 \begin{equation}\label{neo-Hookean}
 W=\frac{\mu}{2}\left(\lambda^2+\lambda^{-2}\lambda_z^{-2}+\lambda_z^{-2}-3\right)+\frac{1}{2\varepsilon}\lambda^{-2}\lambda_z^{-2}D_0^2,
 \end{equation}
 where $\varepsilon$ is the permittivity of the solid, which is independent of the deformation.


\subsection{Static deformation}


In this case Eqs. \eqref{structural-constants}$_2$, \eqref{ab} and \eqref{connection} reduce to
\begin{equation}\label{governing-equation}
\varphi=\frac{\lambda_z\left(\lambda_b^2-\lambda_a^2\right)}{2}\frac{A}{H}, \quad
\lambda_a^2\lambda_b^2\lambda_z^2=\overline D_0^2+1, \quad
\overline V=\frac{2\overline D_0}{\lambda_z^2\left(\lambda_b^2-\lambda_a^2\right)}\text{ln}\frac{\lambda_b}{\lambda_a},
\end{equation} 
where we are using the following non-dimensional measures of voltage and electric vector,
\begin{equation}
\overline V=\frac{V}{H}\sqrt{\frac{\varepsilon}{\mu}}, \quad \overline D_0=\frac{D_0}{\sqrt{\mu\varepsilon}}.
\end{equation} 

For given $\overline V, \varphi, \lambda_z$ and $A/H$, $\lambda_a, \lambda_b$ and $\overline D_0$ can be determined from Eq. \eqref{governing-equation}. Then the dimensionless stresses and electric field in the solid follow from Eqs. \eqref{electric-components}, \eqref{normal-stress}, \eqref{shear-stress} and \eqref{constant-K} as
\begin{equation}
\overline \tau_{rr}(\lambda)=\frac{\tau_{rr}}{\mu}=\frac{\left(\lambda^2-\lambda_a^2\right)\left(\lambda^2-\lambda_b^2\right)}{2\lambda^2}, \quad 
\overline \tau_{\theta\theta}(\lambda)=\frac{\tau_{\theta\theta}}{\mu}=\frac{3\lambda^4-\lambda_a^2\lambda_b^2-\lambda^2\left(\lambda_a^2+\lambda_b^2\right)}{2\lambda^2},
\end{equation} \begin{equation}
\overline E_r=E_r\sqrt{\frac{\varepsilon}{\mu}}=\lambda^{-1}\lambda_z^{-1}\overline D_0.
\end{equation}


\subsubsection{Effect of the voltage}


\begin{figure}[h!]
\centering
\includegraphics[width=0.8\textwidth]{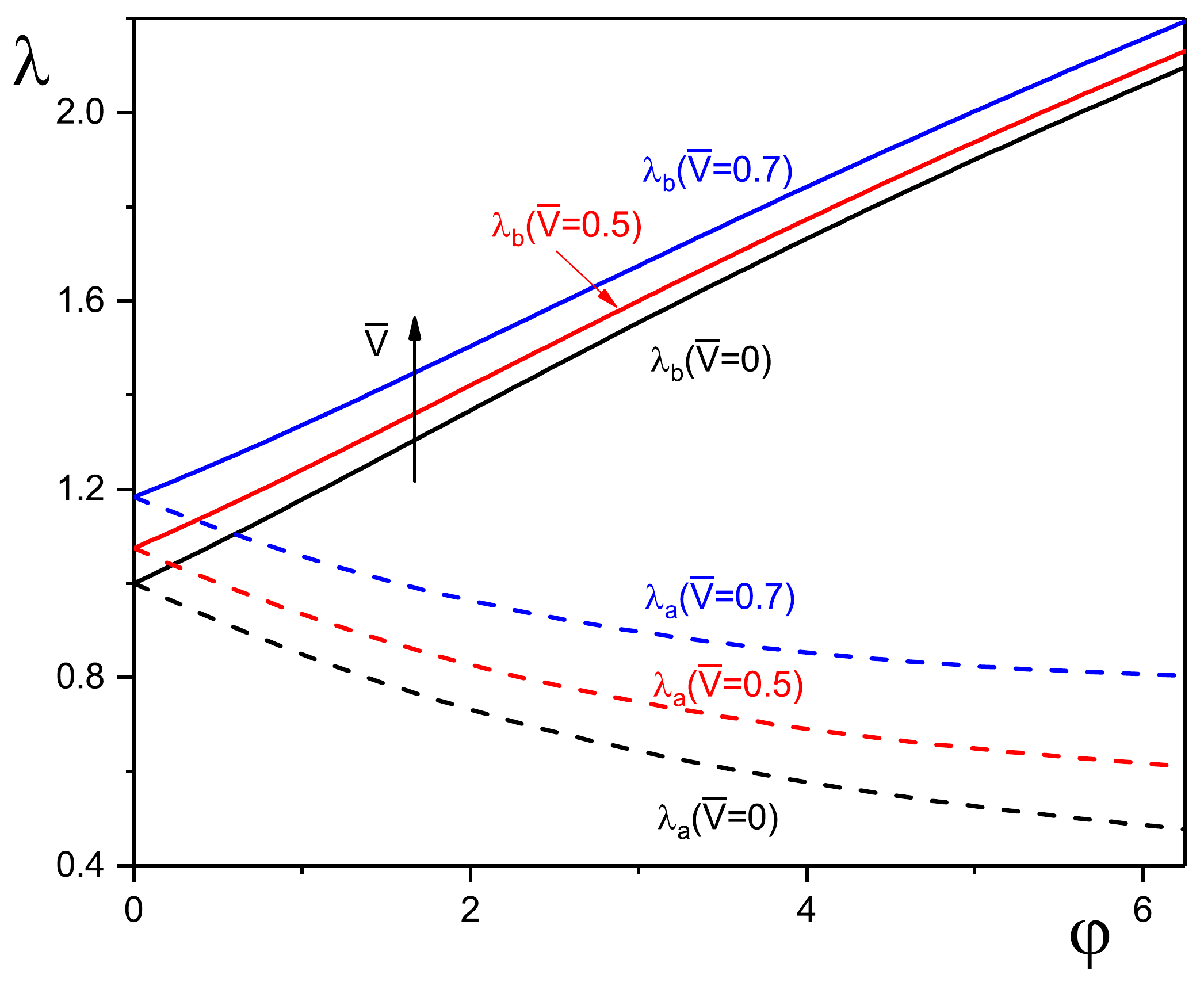}
\caption{
{\footnotesize
Plots of $\lambda_a,\lambda_b-\varphi$ for dielectric slabs with $\lambda_z=1, A/H=3$ subject to voltage $\overline V=0, 0.5, 0.7$, respectively.
}
}
\label{figure3}
\end{figure}

\begin{figure}[h!]
\centering
\includegraphics[width=0.8\textwidth]{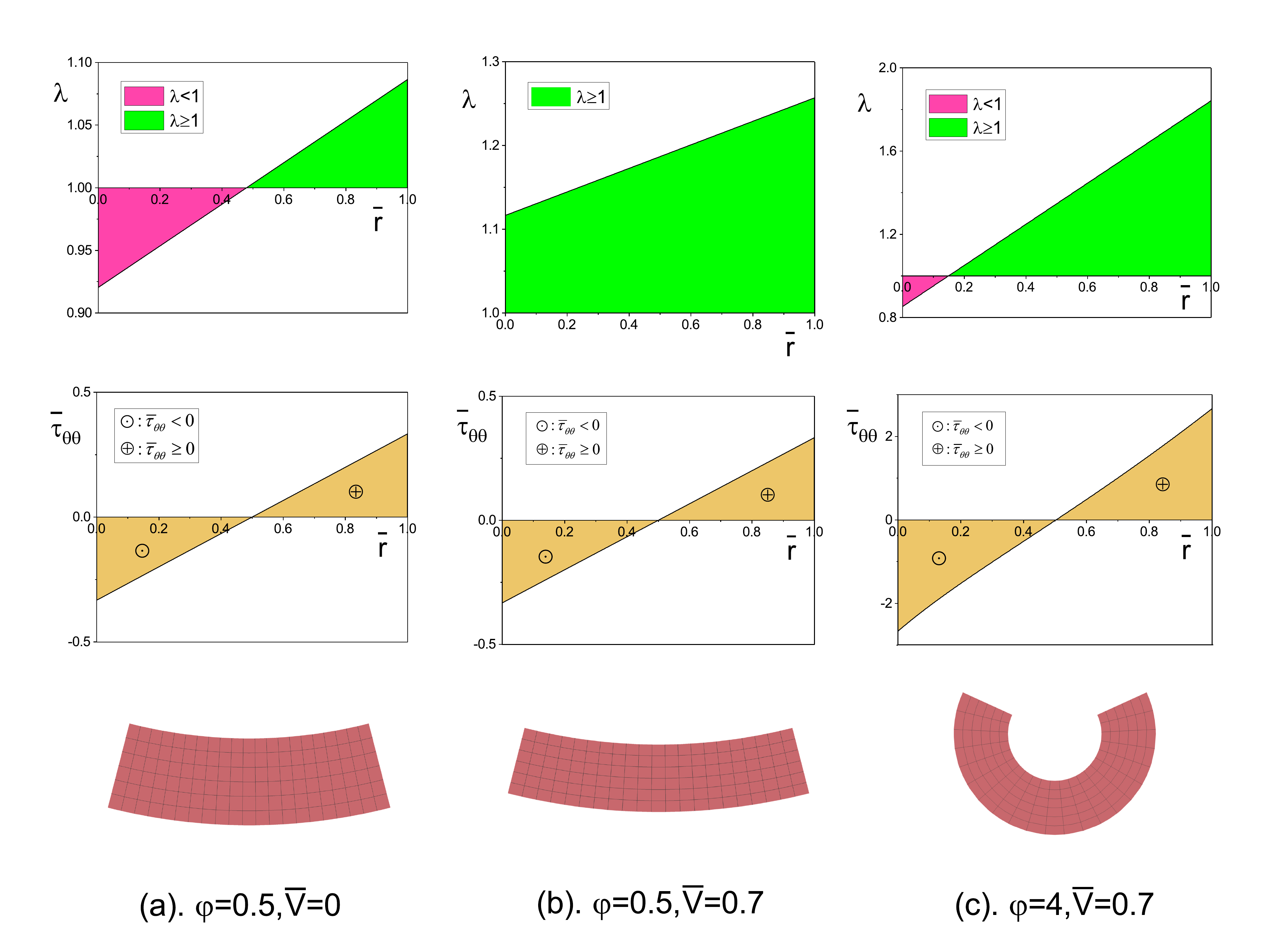}
\caption{
{\footnotesize
Bending of dielectric slabs which are three times wider than thick, with no axial compression ($\lambda_z=1, A/H=3$) and subject to various bending angles and voltage loadings: (a) $\varphi=0.5, \overline V=0$; (b) $\varphi=0.5, \overline V=0.7$; (c) $\varphi=4, \overline V=0.7$. The top, middle and bottom rows correspond to the circumferential stretch, circumferential stress distributions, and bending shapes, respectively.
}
}
\label{figure4}
\end{figure}

\begin{figure}[h!]
\centering
\includegraphics[width=0.8\textwidth]{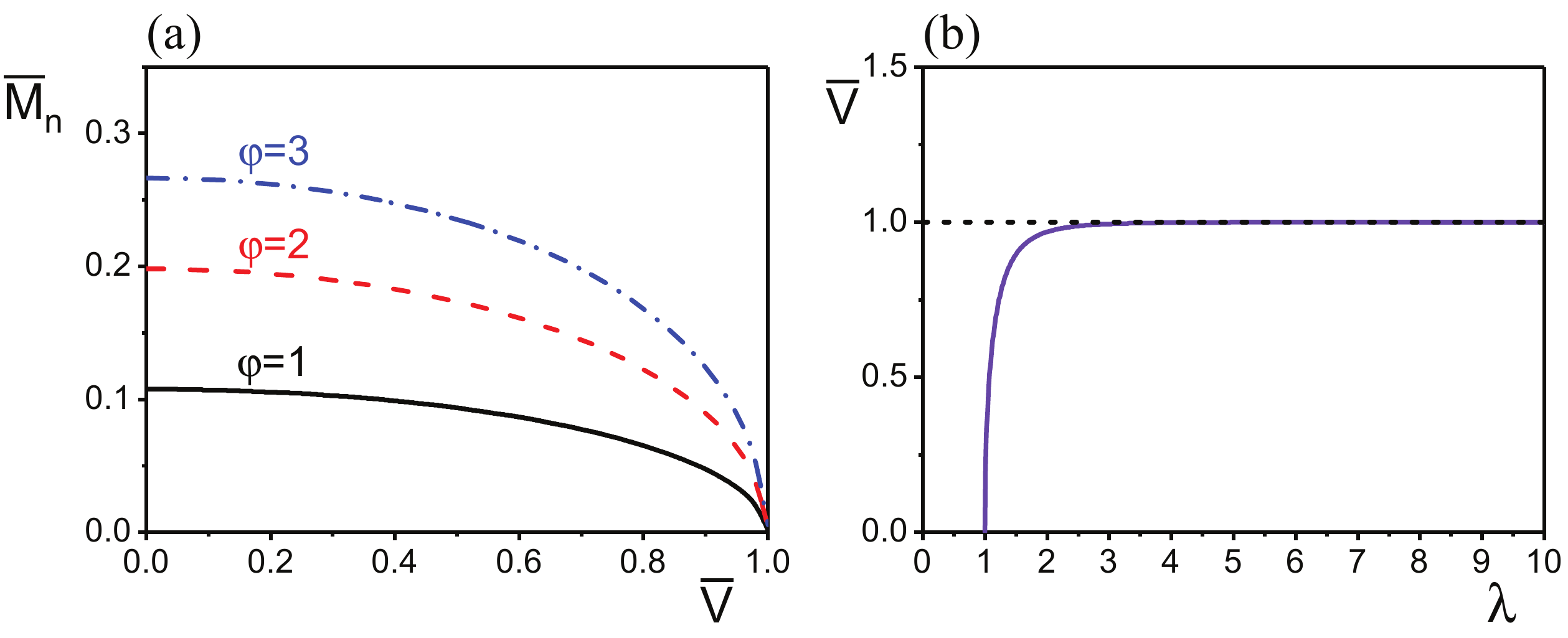}
\caption{
{\footnotesize
\color{black}(a) Plots of $\overline M_n-\overline V$ for several specific bending angles $\varphi=1,2,3$ of a dielectric slab which is three times wider than thick, and no axial compression ($\lambda_z=1, A/H=3$); (b) Nonlinear response of a dielectric slab subject to a voltage.\color{black}
}
}
\label{figure5}
\end{figure}

In Figure \ref{figure3} we plot the circumferential stretches of the bent inner and outer surfaces $\lambda_a$ and $\lambda_b$ versus the bending angle $\varphi$ for different applied voltages $\overline V=0, 0.5, 0.7$, based on Eq. \eqref{governing-equation}. In Figure \ref{figure4}, we plot the distributions of circumferential stretch $\lambda$ and stress $\overline{\tau}_{\theta\theta}$ in the sector and the bending shapes for several given $\varphi$ and $\overline V$. We can see from Eq. \eqref{governing-equation} that the length aspect ratio $L/H$ does not affect the bending deformation of the slab. Here in the calculation the axial constraint and the initial configuration of the slab are fixed as $\lambda_z=1, A/H=3$, and the non-dimensional measure of the radial coordinate $\overline r=\left(r-r_a\right)/\left(r_b-r_a\right)$ is introduced. 

It can be seen from Figure \ref{figure3} that when there is no applied voltage ($\overline V=0$), the slab bends with $\lambda_a$ decreasing and $\lambda_b$ increasing from 1. Hence, the inner face of the sector contracts circumferentially while the outer face stretches (Figure \ref{figure4}$a$), a result which is independent of the value of $\varphi$. With the application of voltage, both $\lambda_a$ and $\lambda_b$  of a slightly bent sector are larger than 1 and hence, every circumferential element in the sector is stretched (Figure \ref{figure4}$b$). If the bending moments are increased, the bending angle increases, and the inner surface eventually contracts circumferentially, and the outer surface is stretched at all times (Figure \ref{figure4}$c$). Note that for a bent sector, $\overline \tau_{\theta\theta}$ depends on $\overline r$ almost linearly, the transverse stress of the inner part of the sector is always compressive while that of the outer part is always tensile, separated by a neutral axis corresponding to $\overline \tau_{\theta\theta}=0$.
 
We learn from Eq. \eqref{normal-force} that only mechanical moments are required to drive the bending of the dielectric slab. 
The effect of the applied voltage $\overline V$ on the moment $\overline M_n$ needed to trigger a specific bending ($\varphi=1,2,3$) of the dielectric slab with $\lambda_z=1, A/H=3$ is presented in \color{black}Figure \ref{figure5}$a$ \color{black}. 
We can see clearly that $\overline M_n$ decreases as $\overline V$ increases, which suggests that the application of the voltage makes the slab easier to be bent. 
\color{black} 
Theoretically, as the applied voltage increases, the dielectric slab thins down, making the slab easier to be bent. 
As a result, the moment needed for the bending decreases. 
For a dielectric slab undergoes plane strain deformation, the maximal electric field $\overline V$ applied cannot exceed the value 1 (Figure \ref{figure5}$b$). 
As the electric field tends to 1, the slab becomes be ultra-thin, and the moment drops to zero (Figure \ref{figure5}$a$). \color{black}


\subsubsection{Effect of the axial compression}


\begin{figure}[h!]
\centering
\includegraphics[width=0.8\textwidth]{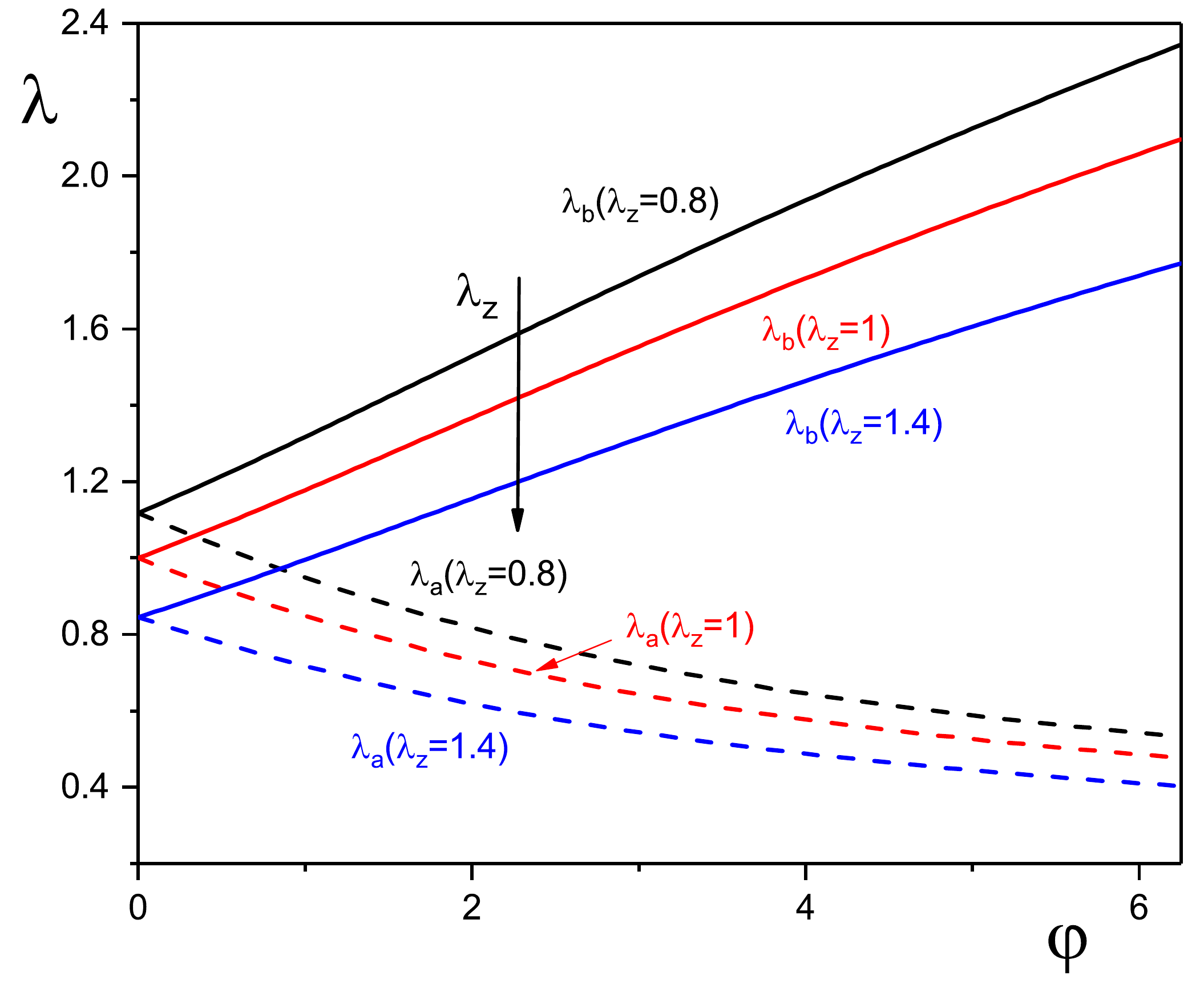}
\caption{
{\footnotesize
Plots of $\lambda_a,\lambda_b-\varphi$ for fixed axial compressions $\lambda_z=0.8,1,1.4$ of dielectric slabs with $\overline V=0, A/H=3$.
}
}
\label{figure6}
\end{figure}

\begin{figure}[h!]
\centering
\includegraphics[width=0.8\textwidth]{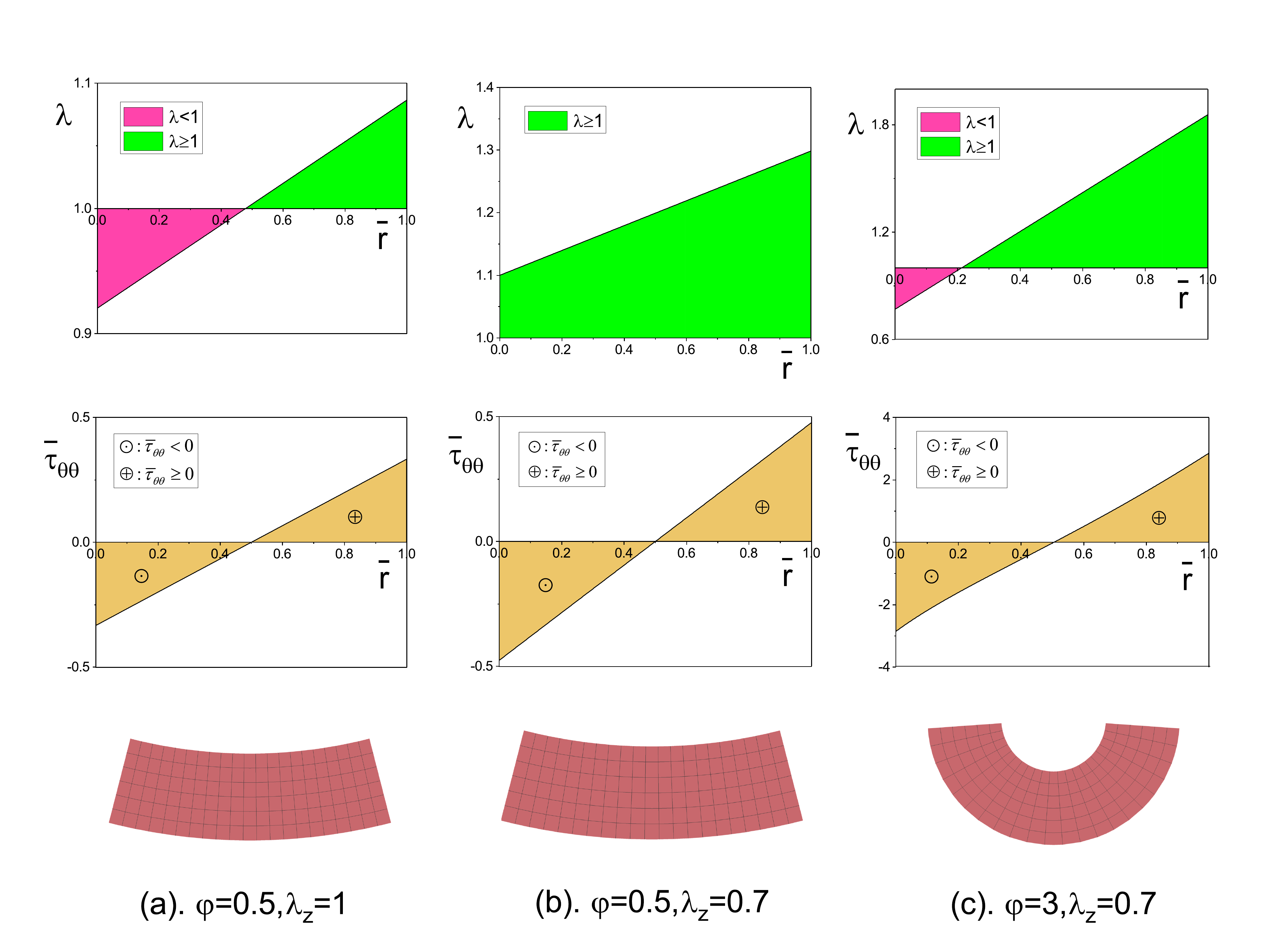}
\caption{
{\footnotesize
Purely elastic bending of a dielectric slab which is three times wider than thick ($\overline V=0, A/H=3$) for various bending angles and axial compression ratios: (a) $\varphi=0.5, \lambda_z=1$; (b) $\varphi=0.5, \lambda_z=0.7$; (c) $\varphi=3, \lambda_z=0.7$. The top, middle and bottom rows show the variations through the thickness of the circumferential stretch and of the stress distributions, and the resulting bending shapes, respectively.
}
}
\label{figure7}
\end{figure}

\begin{figure}[h!]
\centering
\includegraphics[width=0.8\textwidth]{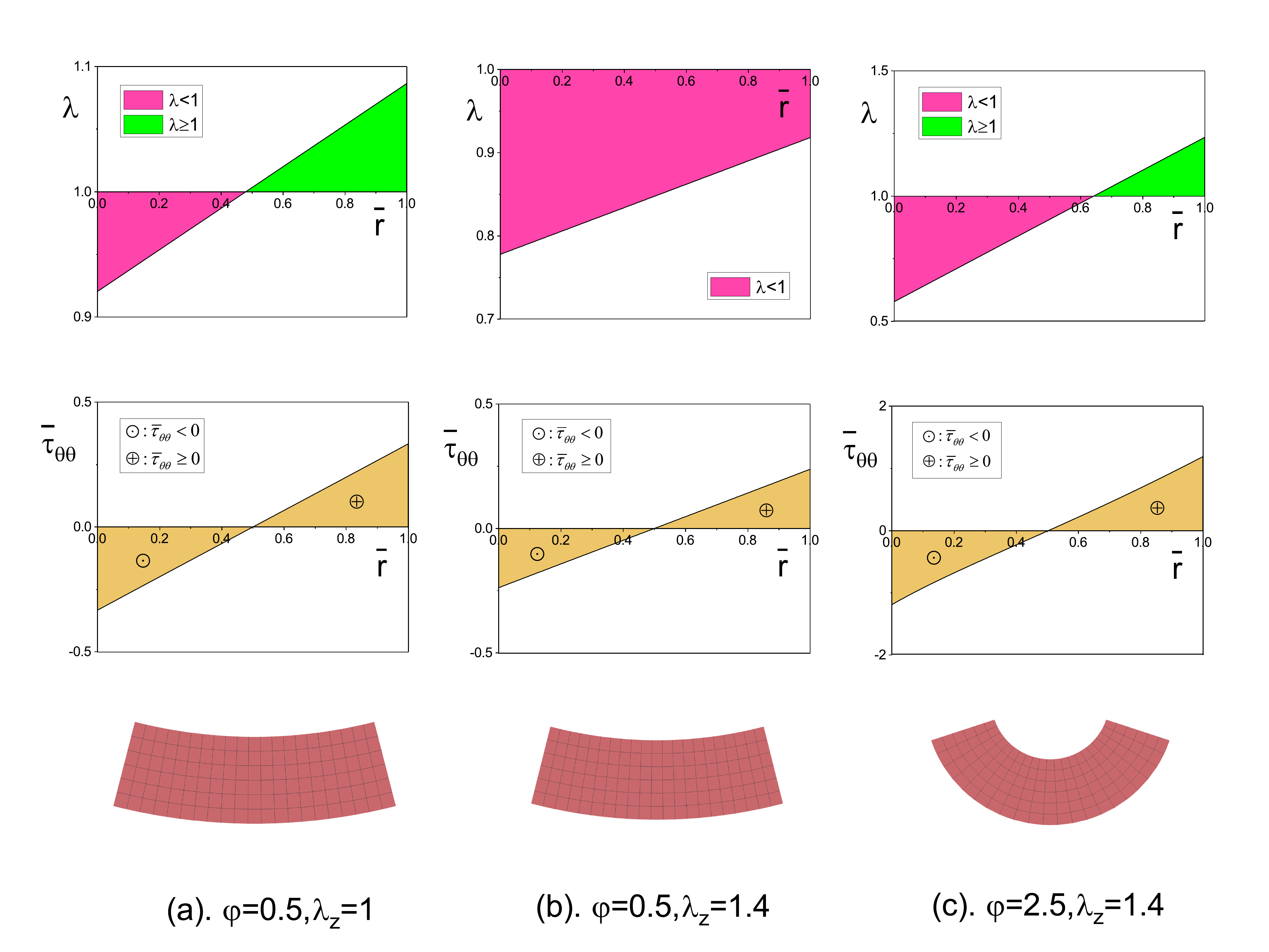}
\caption{
{\footnotesize
Purely elastic bending of a dielectric slab which is three times wider than thick ($\overline V=0, A/H=3$) for various bending angles and axial elongation ratios: (a) $\overline \varphi=0.5, \lambda_z=1$; (b) $\overline \varphi=0.5, \lambda_z=1.4$; (c) $\overline \varphi=2.5, \lambda_z=1.4$. The top, middle and bottom rows show the variations through the thickness of the circumferential stretch and of the stress distributions, and the resulting bending shapes, respectively.
}
}
\label{figure8}
\end{figure}

\begin{figure}[h!]
\centering
\includegraphics[width=0.8\textwidth]{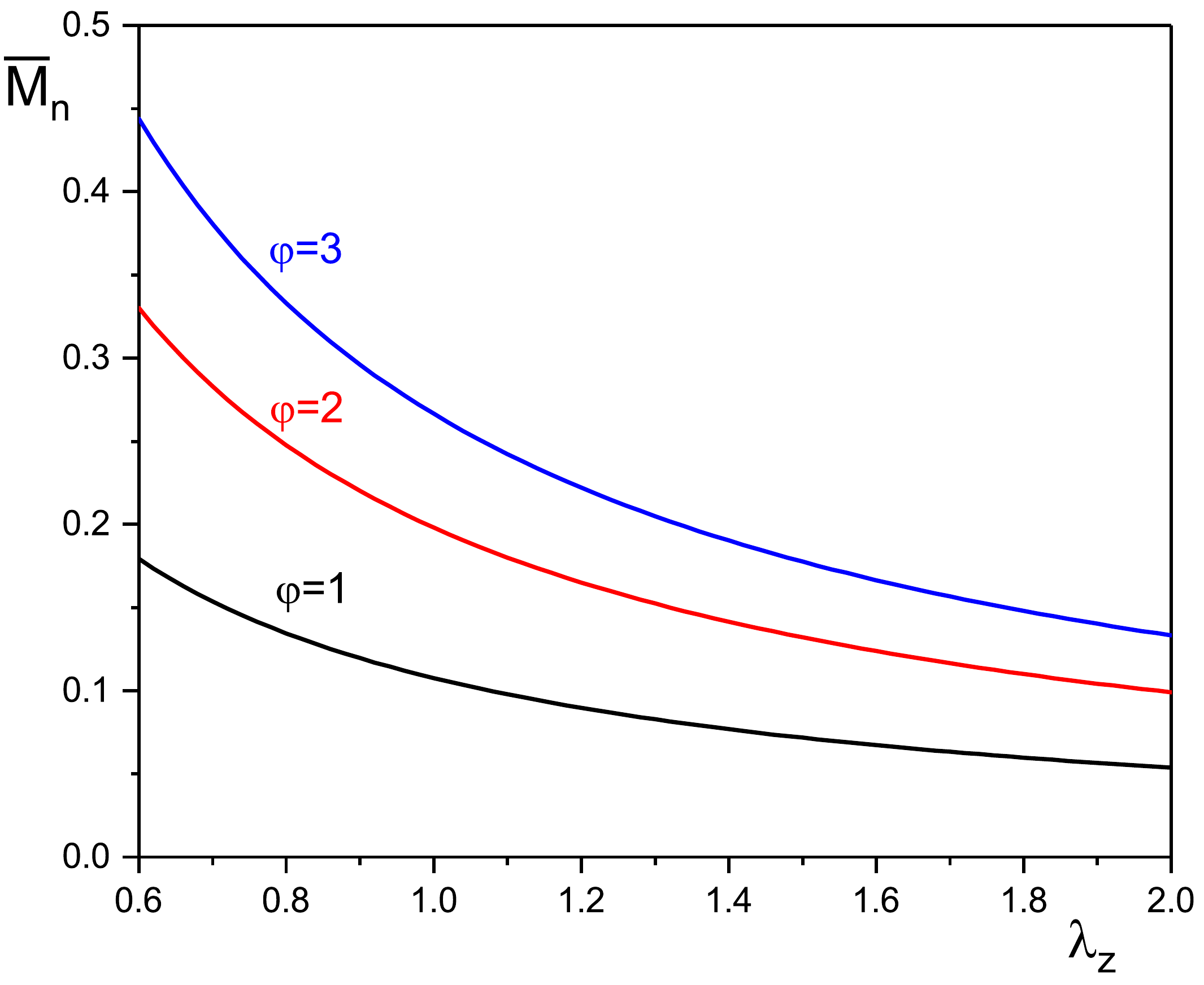}
\caption{
{\footnotesize
Plots of $\overline M_n-\lambda_z$ for bending angle $\varphi=1,2,3$ of a dielectric slab with $\overline V=0, A/H=3$.
}
}
\label{figure9}
\end{figure}

Figures \ref{figure6}-\ref{figure9} illustrate the effect of the axial constraint as measured by the stretch $\lambda_z$ on the finite bending of a dielectric slab with $\overline V=0, A/H=3$. We see that compressive ($\lambda_z<1$) and tensile ($\lambda_z\ge 1$) axial loads produce different effects on the bending deformation (Figure \ref{figure6}). A compressive loading has a similar effect as a voltage $\overline V$ on the bending: when the slab is bent slightly, every circumferential element in the sector is stretched (Figure \ref{figure7}b); as the bending angle $\varphi$ increases to a sufficiently large value, the inner part of the sector contracts circumferentially, and the outer part is stretched (Figure \ref{figure7}c). Conversely, for a pre-stretched, slightly bent slab, every circumferential element of the solid is contracted (Figure \ref{figure8}b); then as $\varphi$ increases, $\lambda_b$ increases, and eventually, the outer part of the solid will be stretched again for a sufficiently large $\varphi$ (Figure \ref{figure8}c). Notice that in both cases, the distribution of circumference stress $\overline {\tau}_{\theta\theta}$ depends almost linearly on $\overline r$. We can see from Figure \ref{figure9} that stretching the slab makes the solid easier to be bent.


\subsection{Stability analysis}


The corresponding material parameters are obtained by substituting Eq. \eqref{neo-Hookean} into Eqs. \eqref{appendix1}-\eqref{appendix3} as
\begin{align}
&A_{01111}=A_{01212}=A_{01313}=\mu\lambda^{-2}\lambda_z^{-2}+D_r^2, && A_{02121}=A_{02222}=A_{02323}=\mu\lambda^{-2}, \notag \\
& A_{03131}=A_{03232}=A_{03333}=\mu\lambda_{z}^{-2}, && \Gamma_{0111}=2\Gamma_{0122}=2\Gamma_{0133}=2D_r, \notag \\
& K_{011}=K_{022}=K_{033}=\varepsilon^{-1}.
\end{align}


\subsubsection{Pure elastic problem}


\begin{figure}[h!]
\centering
\includegraphics[width=0.8\textwidth]{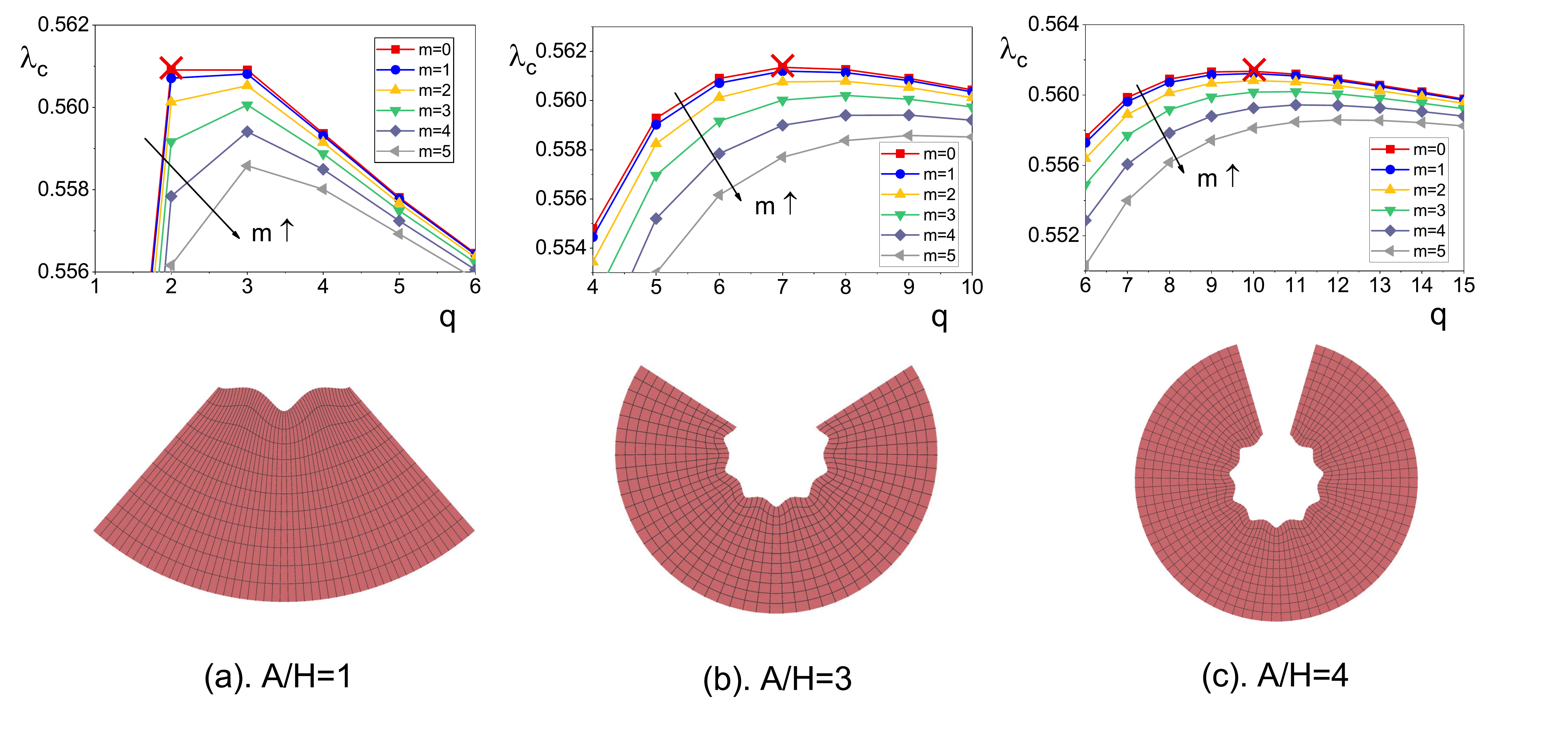}
\caption{
{\footnotesize
Elastic bending instability of a dielectric slab with fixed axial compression ($\overline V=0, \lambda_z=1$) for various width aspect ratios: (a) $A/H=1$; (b) $A/H=3$; (c) $A/H=4$ and fixed length aspect ratio $L/H=10$. The top row shows plots of the critical circumferential stretch $\lambda_c$ versus the number of circumferential wrinkles $q$ for a range of axial modes $m=0-5$. The bottom row shows the corresponding wrinkling shapes when instability occurs. The highest point of $\lambda_c-q$ curves for each ease is marked by cross, representing the onset of the instability. In this case the $m=0$ plot is always on top and there are no axial wrinkles, only circumferential.
}
}
\label{figure10}
\end{figure}

\begin{figure}[h!]
\centering
\includegraphics[width=0.8\textwidth]{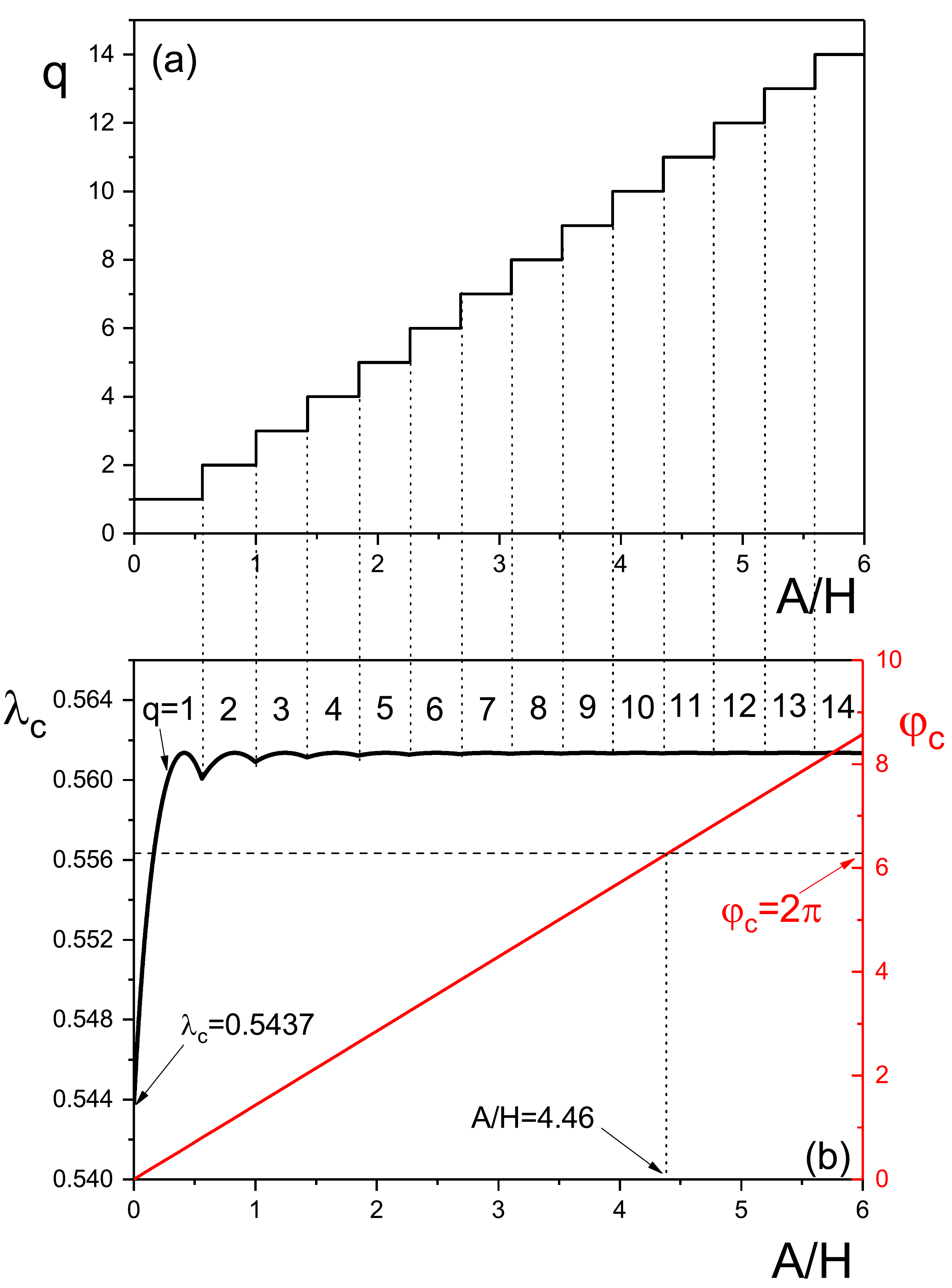}
\caption{
{\footnotesize
(a) Critical circumferential mode number $q$ and (b) stretch $\lambda_c$ (evaluated at the inner face of the bent sector $r_a$) and critical bending angle $\varphi_c$ versus width aspect ratio $A/H$ of an elastic slab ($\overline V=0$) under bending only ($\lambda_z=1$) at the onset of buckling.
}
}
\label{figure11}
\end{figure}

First, we consider the purely elastic slab ($\overline V=0$) under bending only ($\lambda_z=1$), a case which has been previously investigated experimentally \citep{Gent99, Roccabianca10} and theoretically \citep{Triantafyllidis80, Gilchrist09, Destrade09, Roccabianca10}. Figure \ref{figure10} exhibits numerical results for the bending instability for different axial mode numbers $m=0-5$ of elastic slabs with $A/H=1, 3$ and 4, and $L/H=10$, respectively. The solid buckles when the stretch of the inner surface $\lambda_a$ reaches the highest point of the $\lambda_c-q$ curve. We find that the bending instability occurs with decreasing critical stretch $\lambda_c$ as $m$ increases and the buckling mode with $m=0$ always occurs first, indicating that only circumferential wrinkles occur at the onset of instability. For instance, a slab with $A/H=1, 3, 4$ buckles in modes $q=2, 7, 10$ and $m=0$ when the circumferential stretch of the inner surface of the sector reaches $\lambda_c=0.56091, 0.5614, 0.56135$ and the bending angle reaches $\varphi_c=1.43, 4.23, 5.72$, respectively. Notice that the perturbation decays dramatically along the radius, and that the displacement on the inner face is several orders of magnitude larger than that on the outer face.

Figure \ref{figure11} reports the critical number of circumferential wrinkles $q$, the critical stretch $\lambda_c$ and the critical bending angle $\varphi_c$ as functions of the aspect ratio $A/H$. For a given $A/H$, each mode number $q$ corresponds a different value of the critical stretch $\lambda_c$ and a series of branches can be obtained by taking $q=1,2,3,...$. However only the highest value is meaningful, thus the other curves below the highest curve are not presented in the $\lambda_c-A/H$ plot. We observe that as $A/H$ increases, the mode number $q$ increases, indicating that more wrinkles appear as instability occurs for a more slender slab. The critical bending angle $\varphi_c$ increases linearly as $A/H$ increases. For a slab with sufficiently large width aspect ratio $A/H$($>4.46$), the structure can be bent into a tube without encountering any instability (Figure \ref{figure11}b). In the half-space limit ($A/H \to 0$), the critical stretch is $\lambda_c=0.5437$, which corresponds to the threshold value of surface instability of a compressed elastic slab \citep{Biot65, Destrade09}. When $A/H$ is small, the critical stretch $\lambda_c$ varies significantly as $A/H$ varies. While for slab with sufficiently large $A/H$, $\lambda_c$ reaches a horizontal asymptote $\lambda_c \approx 0.5618$.
\subsubsection{Effect of the voltage}
\begin{figure}[h!]
\centering
\includegraphics[width=0.8\textwidth]{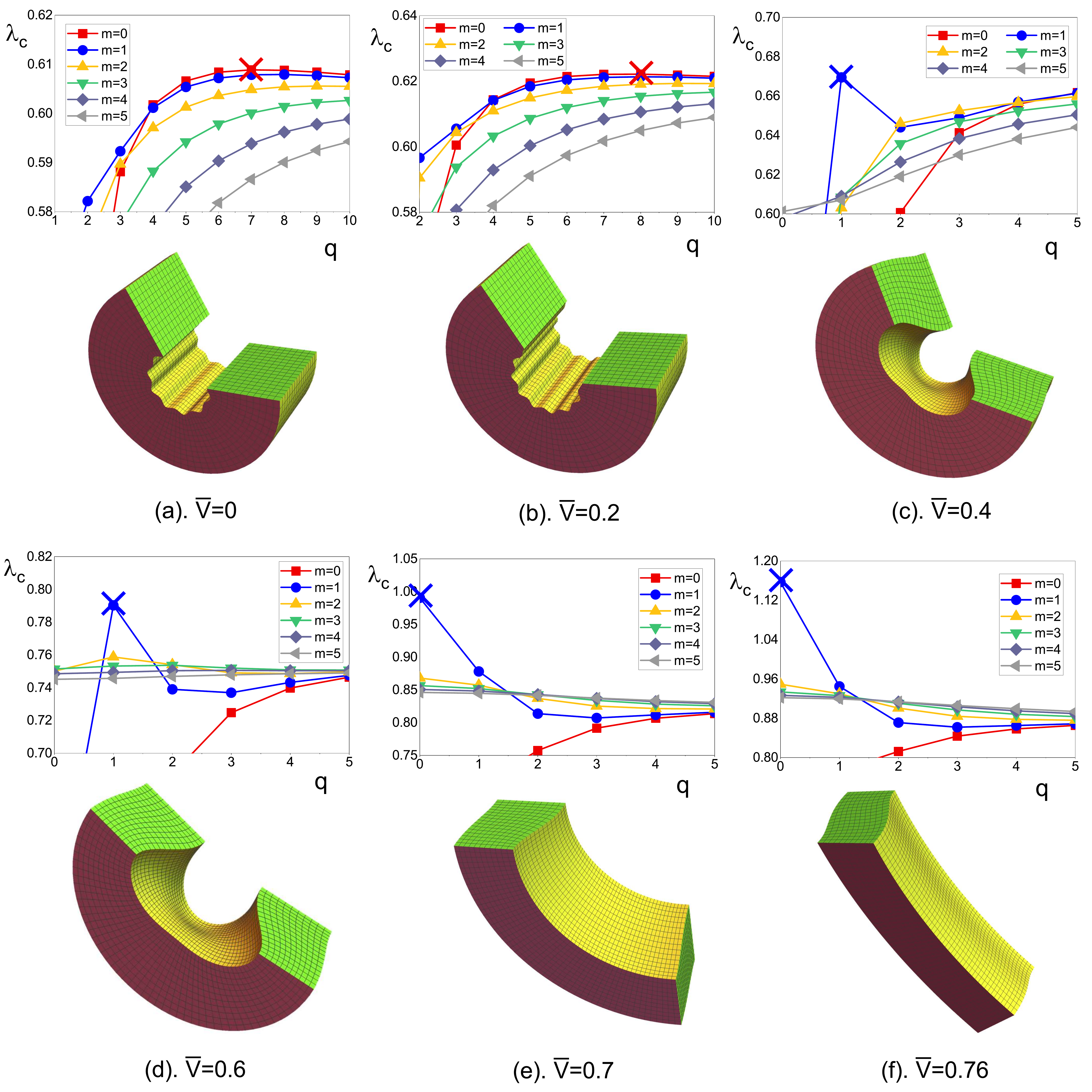}
\caption{
{\footnotesize
Bending instability of compressed dielectric slabs which are 3 times wider and 1.5 times taller than thick ($A/H=3, L/H=1.5, \lambda_z=0.85$) and subject to increasing voltages: (a) $\overline V=0$; (b) $\overline V=0.2$; (c) $\overline V=0.4$; (d) $\overline V=0.6$; (e) $\overline V=0.7$; (f) $\overline V=0.76$. The top rows are plots of $\lambda_c$ versus $q$ for a range of modes $m=0-5$ and bottom rows are the corresponding wrinkling shapes when instability occurs. In cases (a) and (b), circumferential wrinkles occurs and in cases (e) and (f), axial wrinkles occurs whereas in cases (c) and (d), a two-dimensional (circumferential and axial) pattern emerges.
}
}
\label{figure12}
\end{figure}

\begin{figure}[h!]
\centering
\includegraphics[width=0.8\textwidth]{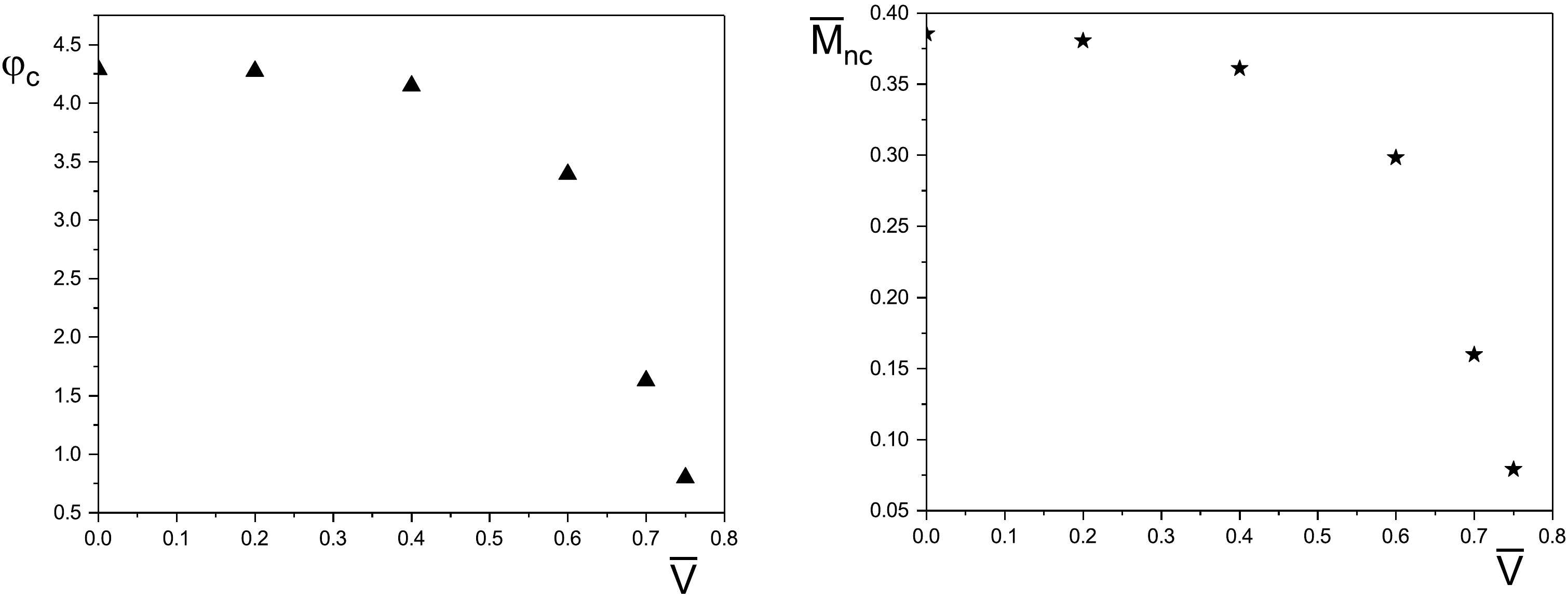}
\caption{
{\footnotesize
Effect of the applied voltage $\overline V$ on the critical values of bending angle $\varphi_c$ and moment $\overline M_{nc}$ for dielectric slabs with $A/H=3, L/H=1.5, \lambda_z=0.85$.
}
}
\label{figure13}
\end{figure}

We now consider the effect of the applied voltage $\overline V$ on the bending instability of a dielectric slab. Figure \ref{figure12} presents plots of $\lambda_c$ versus $q$ for a range of modes $m=0-5$ and the corresponding wrinkling shapes when instability occurs for dielectric slabs with $A/H=3, L/H=1.5$ and subject to $\overline V=0,0.2,0.4,0.6,0.7,0.76$. Here we fix the axial deformation of the bending deformation as a 15\% contraction ($\lambda_z=0.85$). For each of the cases (a)-(f) shown in Figure \ref{figure12}, the buckling mode is $(m,q)=(0,7), (0,8),(1,1),(1,1),(1,0)$ and (1,0), respectively. The critical stretch $\lambda_c$ increases as $\overline V$ increases. For the cases where the applied voltage is small, only circumferential wrinkles occur $(m=0, q\neq 0)$ when bending buckling happens, and the mode number $q$ increases as the voltage increases (Figures \ref{figure12}a, b). As the voltage increases further, both circumferential and axial wrinkles occur simultaneously $(m\neq 0, q\neq 0)$ at the onset of bending instability (Figures \ref{figure12}c, d) and combine to give a 2D pattern. Finally, for dielectric slabs subject to sufficiently large voltage, a slight bending will drive the instability of the structure and in this case, only axial wrinkle occurs ($(m\neq 0, q=0)$, see Figure \ref{figure12}e, f). It should be mentioned that the maximal number of axial wrinkle is one ($m=0, 1$).

We extract the critical bending angle $\varphi_c$ when the instability occurs and the critical moment $\overline M_{nc}$ needed to drive the instability for the cases presented in Figure \ref{figure12}, and plot them in Figure \ref{figure13} as the applied voltage $\overline V$ changes. It can be seen that both $\overline \varphi_c$ and $\overline M_{nc}$ decrease as $\overline V$ increases, indicating that the application of the voltage makes the dielectric slab more susceptible to fail. One may expect that $\overline \varphi_c$ and $\overline M_{nc}$ will be zero for a critical $\lambda_{zc}$, corresponding the critical value of instability of a compressed elastic slab \citep{Dorf14a, Biot63}.
\subsubsection{Effect of the axial constraint}
\begin{figure}[h!]
\centering
\includegraphics[width=0.8\textwidth]{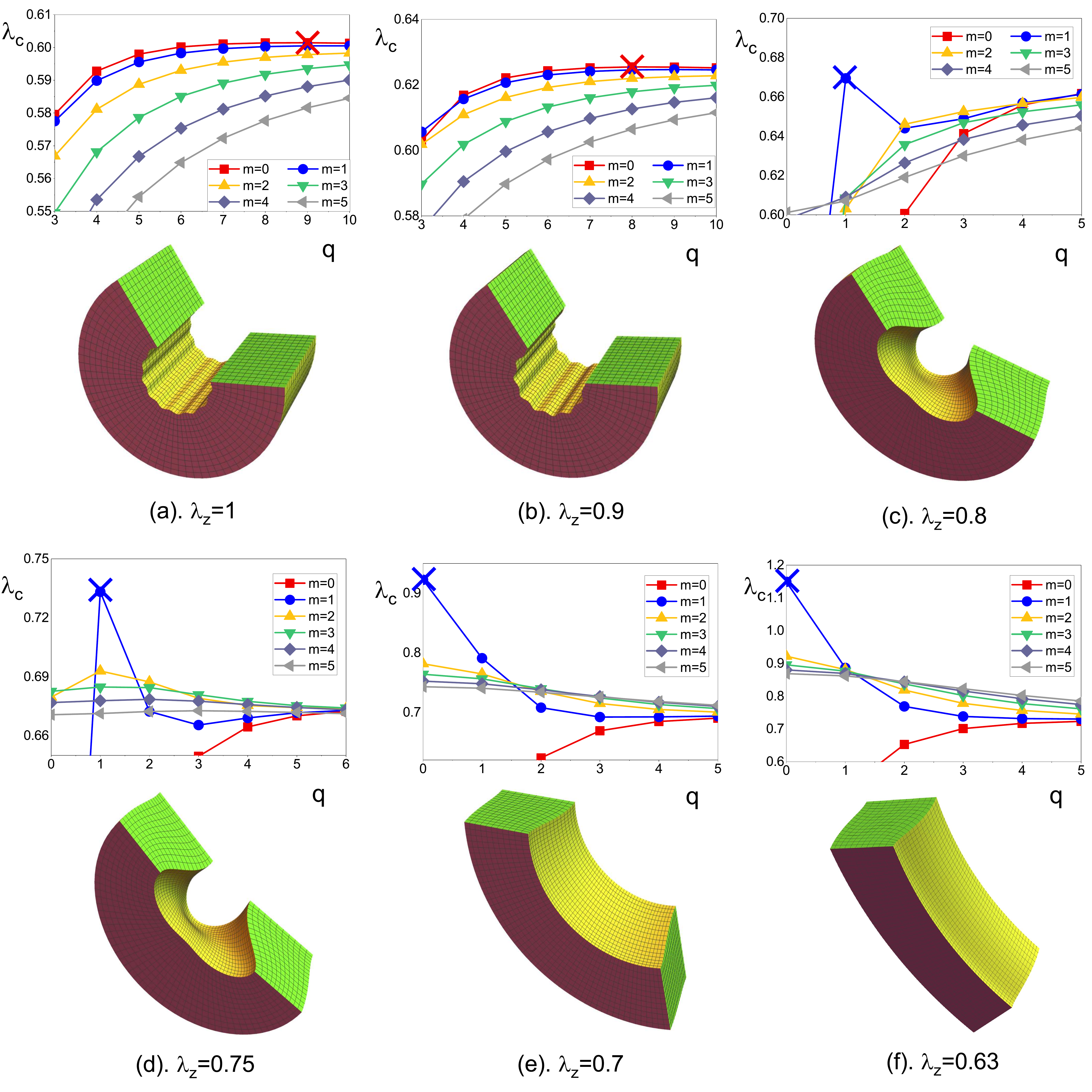}
\caption{
{\footnotesize
Bending instability of dielectric slabs with $A/H=3, L/H=1.5, \overline V=0.3$ and subject to (a) $\lambda_z=1$; (b) $\lambda_z=0.9$; (c) $\lambda_z=0.8$; (d) $\lambda_z=0.75$; (e) $\lambda_z=0.7$; (f) $\lambda_z=0.63$. The top rows are plots of $\lambda_c$ versus $q$ for a range of modes $m=0-5$ and bottom rows are the corresponding wrinkling shapes when instability occurs. In cases (a) and (b), circumferential wrinkles occurs and in cases (e) and (f), axial wrinkles occurs whereas in cases (c) and (d), a two dimensional (circumferential and axial) pattern emerges.
}
}
\label{figure14}
\end{figure}

\begin{figure}[h!]
\centering
\includegraphics[width=0.8\textwidth]{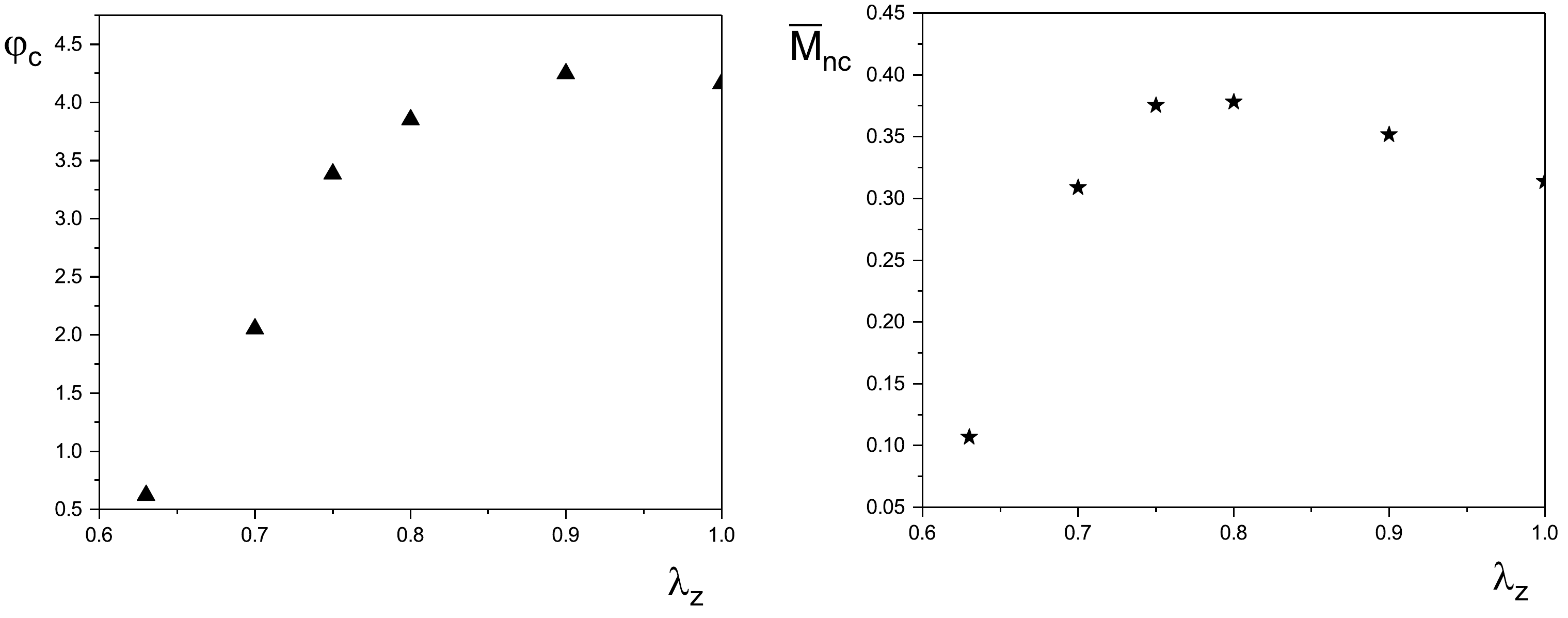}
\caption{
{\footnotesize
Effect of the axial compression, as measured by the axial stretch ratio $\lambda_z$, on the critical values of bending angle $\varphi_c$ and moment $\overline M_{nc}$ for dielectric slabs with $A/H=3, L/H=1.5, \overline V=0.3$.
}
}
\label{figure15}
\end{figure}

Here, we investigate the effect of the axial compression, as measured by the axial stretch ratio $\lambda_z$, on the bending instability of dielectric slabs. 
Figure \ref{figure14} displays numerical results for the bending instability of dielectric slabs with $A/H=3, L/H=1.5$ and subject to $\overline V=0.3$ and $\lambda_z=1,0.9,0.8,0.75,0.7,0.63$, respectively. 
Figure \ref{figure15} presents the corresponding $\varphi_c$ and $\overline M_{nc}$ when buckling occurs. 
For each of the cases (a)-(f) shown in Figure \ref{figure14}, the buckling mode is $(m,q)=(0,9), (0,8),(1,1),(1,1),(1,0)$ and (1,0), respectively. 
We can see that decreasing the axial stretch ratio  has a similar effect as increasing the applied voltage $\overline V$ on the bending buckling behavior of dielectric slabs, i.e., the critical stretch $\lambda_c$ increases as $\lambda_z$ decreases, and circumferential wrinkles occur first and eventually only axial wrinkles exist as $\lambda_z$ decreases to a sufficiently small value. 
Note that when the axial compression is small, the mode number $q$ decreases as $\lambda_z$ increases (Figures \ref{figure14}a, b), which is different from the case of increasing $\overline V$ (Figures \ref{figure12}a, b). 
Due to the competition mechanisms of the effects of $\overline V$ and $\lambda_z$ on the bending instability of the structure, the $\varphi_c, \overline M_{nc}-\lambda_z$ curves are non-monotone.
On the one hand, decreasing $\lambda_z$ increases the thickness of the slab and thus decreases the true electric field, which consequently increases the stability of the structure. 
On the other hand, decreasing $\lambda_z$ makes the structure be easier to fail in the axial direction and poses a destabilizing influence on the slab. As a result, $\varphi_c$ and $\overline M_{nc}$ increase first and then decrease to zero, as $\lambda_z$ decreases (Figure \ref{figure15}), indicating that the voltage $\overline V$ plays a major role when the structure is only slightly compressed, while the axial compression $\lambda_z$ presents the dominant influence when the structure is dramatically compressed.


\section{Conclusions}
\label{section5}



We presented a theoretical analysis of the finite bending deformation and the associated bending instability of an incompressible dielectric slab subject to a combined action of voltage and mechanical moments. 
We derived the three-dimensional equations governing the static finite bending deformation and the associated incremental deformation of the slab for a general form of energy function. 
In particular, we studied explicit expressions of the radially inhomogeneous biasing fields in the slab for ideal neo-Hookean dielectric materials. 
We took the electric loading to be voltage-controlled and so we chose a state vector accordingly to rewrite the incremental governing equation in the Stroh differential form. 
We used the surface impedance matrix method to obtain numerically the bending threshold for the onset of the instability and the wrinkled shape of the shell when bending instability occurs.

We first studied the effects of the applied voltage and axial compression on the finite bending deformation.
We showed that the length aspect ratios of the slab $L/H$ does not affect the bending deformation of the slab. 
The applied voltage increases the circumferential stretch in the body so that every circumferential element in a slightly bent slab is stretched. 
The moments needed to drive the specific bending of the slab decrease as the voltage increases, indicating that the application of the voltage makes the slab easier to bend. 
We found that the compressive axial constraint has a similar effect as the applied voltage, while on the contrary, every circumferential element in a slightly bent slab, subject to axial pre-stretch, is contracted. 
As the axial stretch increases, the moments needed to drive a specific bending of the slab decrease, indicating that the axial pre-stretch makes the slab easier to bend. 
In any case, the circumferential stretch deforms linearly along the radial direction and the transverse stress of the inner part of the sector is always compressive while that of the outer part is always tensile.

We then investigated the combined influences of the applied voltage and axial constraint on the instability of a dielectric slab. 
We obtained the critical circumferential stretch on the inner surface of the deformed shell, as well as the wrinkled shape when the bending instability occurs. 
We recovered the results of the purely elastic problem to validate our analysis. 
Theoretically, the application of the voltage and the axial constraint both play a destabilizing effect, i.e., make the slab more susceptible to wrinkling instability. 
The two effects compete with each other, and an increase in the axial compressive loads leads to a decrease in the true electric field in the body. The applied voltage plays the main role when the constraint is small, while the constraint becomes dominant when the compression is sufficiently large.

In this article we focused on the formation of small-amplitude wrinkles in a bent and axially compressed dielectric slab. 
We did not look at post-buckling behavior or if creases might have preceded wrinkles. 
This is certainly the case in the \emph{in-plane compression of an elastic half-space}, where creases form much earlier ($\lambda_c=0.65$) than the wrinkles predicted by the linearised buckling analysis of Biot ($\lambda_c=0.54$), i.e. with more than 10\% strain difference \citep{Hong09}.
However, recent Finite Element simulations show \color{black} that \color{black} in \emph{bending}, creases occur only a few percent of strain earlier than  wrinkles,  and that their number and wavelength can be predicted by the linearized analysis \citep{Sigaeva18}. 
Hence \color{black} we argue that our analysis is justified as a good approximation for predicting the onset and wavelength of buckling, although of course a fully multi-physics Finite Element Analysis is required to settle this question. 
\color{black}
Moreover, wrinkles have indeed been observed in loaded dielectric elastomers with free sides (e.g. \cite{DuPl06, Liu16}).
\color{black}
To create creases in a dielectric membrane, one could glue one side of a slab to a rigid, conducting substrate, as done by \cite{Wang13}, but the corresponding boundary value problem is then different from the one studied here, where both sides were free of traction.  \color{black}


\section*{Acknowledgments}


This work was supported by a Government of Ireland Postdoctoral  Fellowship from the Irish Research Council and by the National Natural Science Foundation of China (No. 11621062).
MD thanks Zhejiang University for funding a research visit to Hangzhou.


\section*{References}







\appendix
\section{Non-zero electro-elastic moduli}


Here we use the incremental theory of electro-elasticity to compute the non-zero components of the instantaneous electro-elastic moduli with respect to the specific deformation gradient \eqref{deformation-gradient}, as follows \citep{Wu17, Dorf10a}
\begin{align}\label{appendix1}
A_{01111}=& 2\lambda^{-4}\lambda_z^{-4}\left\{\lambda^4\left[2\Omega_{22}+\lambda_z^2\left(\Omega_2+4\Omega_{25}D_r^2\right)+\lambda_z^4D_r^2\left(\Omega_5+2\Omega_{55}D_r^2\right)\right] \right. \notag \\
& \left. +2\left[\Omega_{11}+\lambda_z^4\Omega_{22}+2\lambda_z^2\left(\Omega_{12}+2\Omega_{26}D_r^2\right)+4D_r^2\left(\Omega_{16}+\Omega_{66}D_r^2\right)\right]\right. \notag \\
& \left. +\lambda^2\lambda_z^4\left(\Omega_{2}+4\Omega_{25}D_r^2\right)+4\lambda^2\left(\Omega_{12}+2\Omega_{26}D_r^2\right)\right. \notag \\
& \left. +\lambda^2\lambda_z^2\left[\Omega_{1}+4\Omega_{22}+8\Omega_{56}D_r^4+D_r^2\left(4\Omega_{15}+6\Omega_{6}\right)\right]
\right\}, \notag \\
A_{01122}=& 4\lambda^{-2}\lambda_z^{-4}\left\{\Omega_{12}+\lambda_z^2\Omega_{22}+\lambda^4\lambda_z^2\left[\Omega_{12}+\lambda_z^2\Omega_{22}+\lambda_z^2D_r^2\left(\Omega_{15}+\lambda_z^2\Omega_{25}\right)\right]+2\Omega_{26}D_r^2 \right. \notag \\
& \left. +\lambda^2\left[\Omega_{22}+\lambda_z^6\Omega_{22}+\lambda_z^2\Omega_{11}+\lambda_z^2\Omega_{2}+\lambda_z^2D_r^2\left(2\Omega_{16}+\Omega_{25}\right)+2\lambda_z^4\left(\Omega_{12}+\Omega_{26}D_r^2\right)\right]
\right\}, \notag \\
A_{01133}=& 4\lambda^{-4}\lambda_z^{-2}\left\{\Omega_{12}+\lambda^2\Omega_{22}+\lambda^2\lambda_z^4\left[\Omega_{12}+\lambda^2\Omega_{22}+\lambda^2D_r^2\left(\Omega_{15}+\lambda^2\Omega_{25}\right)\right]+2\Omega_{26}D_r^2 \right. \notag \\
& \left. +\lambda_z^2\left[\Omega_{22}+\lambda^6\Omega_{22}+\lambda^2\Omega_{11}+\lambda^2\Omega_{2}+\lambda^2D_r^2\left(2\Omega_{16}+\Omega_{25}\right)+2\lambda^4\left(\Omega_{12}+\Omega_{26}D_r^2\right)\right]
\right\}, \notag \\
A_{01212}=& 2\lambda^{-2}\lambda_z^{-2}\left\{\Omega_1+2\Omega_6D_r^2+\lambda_z^2\left[\Omega_2+\lambda^2D_r^2\left(\Omega_5+\lambda^2\Omega_6\right)\right]\right\}, \notag \\
A_{01313}=& 2\lambda^{-2}\lambda_z^{-2}\left\{\Omega_1+2\Omega_6D_r^2+\lambda^2\left[\Omega_2+\lambda_z^2D_r^2\left(\Omega_5+\lambda_z^2\Omega_6\right)\right]\right\}, \notag \\
A_{01221}=&-2\lambda_z^{-2}\Omega_2+2\lambda^{2}\Omega_6D_r^2, \quad
A_{01331}=-2\lambda^{-2}\Omega_2+2\lambda_z^{2}\Omega_6D_r^2, \notag \\
A_{02121}=& 2\lambda^2\left(\Omega_1+\lambda_z^2\Omega_2+\Omega_6D_r^2\right), \quad
A_{03131}= 2\lambda_z^2\left(\Omega_1+\lambda^2\Omega_2+\Omega_6D_r^2\right), \notag \\
A_{02222}=& 2\lambda_z^{-4}\left[\lambda_z^2\Omega_2+2\Omega_{22}+\lambda^2\left(\lambda_z^4\Omega_1+4\lambda_z^4\Omega_{12}+\lambda_z^6\Omega_2+4\lambda_z^4\Omega_{22}\right) \right. \notag \\ 
& \left.+2\lambda^4\lambda_z^4\left(\Omega_{11}+2\lambda_z^2\Omega_{12}+\lambda_z^4\Omega_{22}\right)\right], \notag \\
A_{02233}=& 4\lambda^{-2}\lambda_z^{-2}\left[\Omega_{22}+\lambda^4\lambda_z^2\left(\lambda_z^4\Omega_{12}+\lambda_z^2\Omega_{11}+\lambda_z^2\Omega_2+\Omega_{22}\right)\right. \notag \\
&\left. +\lambda^6\lambda_z^4\left(\Omega_{12}+\lambda_z^2\Omega_{22}\right)+\lambda^2\left(2\lambda_z^2\Omega_{12}+\lambda_z^4\Omega_{22}\right)\right], \notag \\
A_{02323}=&2\lambda^2\Omega_1+2\lambda_z^{-2}\Omega_2, \quad A_{02332}=-2\lambda^2\lambda_z^{2}\Omega_2, \quad A_{03232}=2\lambda_z^2\Omega_1+\lambda^{-2}\Omega_2, \notag \\
A_{03333}=& 2\lambda^{-4}\left[\lambda^2\Omega_2+2\Omega_{22}+\lambda_z^2\left(\lambda^4\Omega_1+4\lambda^4\Omega_{12}+\lambda^6\Omega_2+4\lambda^4\Omega_{22}\right) \right. \notag \\ 
& \left.+2\lambda^4\lambda_z^4\left(\Omega_{11}+2\lambda^2\Omega_{12}+\lambda^4\Omega_{22}\right)\right], 
\end{align}
\begin{align}
\Gamma_{0111}=& 4\lambda^{-4}\lambda_z^{-4}D_r\left\{\Omega_{16}+\lambda_z^2\Omega_{26}+\lambda^6\lambda_z^4\left(\Omega_{24}+\lambda_z^2\Omega_{45}D_r^2\right) \right. \notag \\
&\left. +\lambda^4\lambda_z^z\left[\lambda_z^4\Omega_{24}+\Omega_{25}+\lambda_z^2\left(\Omega_{14}+\Omega_{5}+2\Omega_{46}D_r^2+\Omega_{55}D_r^2\right)\right]\right. \notag \\
&\left. +2\Omega_{66}D_r^2+\lambda^2\left[\lambda_z^4\Omega_{25}+\Omega_{26}+\lambda_z^2\left(\Omega_{15}+3\Omega_{56}D_r^2+2\Omega_6\right)\right]
\right\}, \notag \\
\Gamma_{0122}=& 2\lambda^{-2}\lambda_z^{-2}D_r\left[\lambda^2\lambda_z^2\Omega_5+\left(1+\lambda^4\lambda_z^2\right)\Omega_6\right], \notag \\
\Gamma_{0133}=&2\lambda^{-2}\lambda_z^{-2}D_r\left[\lambda^2\lambda_z^2\Omega_5+\left(1+\lambda^2\lambda_z^4\right)\Omega_6\right], \notag \\
\Gamma_{0221}=& 4\lambda^{-2}\lambda_z^{-4}D_r\left[\lambda^6\lambda_z^6\left(\Omega_{14}+\lambda_z^2\Omega_{24}\right)+\lambda^4\lambda_z^4\left(\Omega_{15}+\Omega_{24}+\lambda_z^2\Omega_{25}\right) \right. \notag \\
& \left. +\Omega_{26}+\lambda^2\lambda_z^2\left(\Omega_{16}+\Omega_{25}+\lambda_z^2\Omega_{26}\right)\right], \notag \\
\Gamma_{0331}=& 4\lambda^{-4}\lambda_z^{-2}D_r\left[\lambda^6\lambda_z^6\left(\Omega_{14}+\lambda^2\Omega_{24}\right)+\lambda^4\lambda_z^4\left(\Omega_{15}+\Omega_{24}+\lambda^2\Omega_{25}\right) \right. \notag \\
& \left. +\Omega_{26}+\lambda^2\lambda_z^2\left(\Omega_{16}+\Omega_{25}+\lambda^2\Omega_{26}\right)\right], 
\end{align}
\begin{align}\label{appendix3}
K_{011}=& \lambda^{-4}\lambda_z^{-4}\left[2\lambda^2\lambda_z^2\left(\lambda^2\lambda_z^2\Omega_5+\Omega_{6}+\lambda^4\lambda_z^4\Omega_4\right)\right. \notag \\ 
& \left. +4D_r^2\left(\lambda^8\lambda_z^8\Omega_{44}+2\lambda^6\lambda_z^6\Omega_{45}+2\lambda^4\lambda_z^4\Omega_{46}+\lambda^4\lambda_z^4\Omega_{55}+2\lambda^2\lambda_z^2\Omega_{56}+\Omega_{66}\right)\right], \notag \\
K_{022}=&2\left(\Omega_5+\lambda^2\Omega_6+\lambda^{-2}\Omega_4\right), \quad 
K_{033}=2\left(\Omega_5+\lambda_z^2\Omega_6+\lambda_z^{-2}\Omega_4\right).
\end{align}


\section{Derivation of the Stroh formulation}


First, rewriting Eq. \eqref{incremental-incompressibility} by using solutions \eqref{incremental-solutions} gives
\begin{equation}\label{line1}
U_r'=-\frac{1}{r}\left(U_r+nU_\theta+krU_z\right).
\end{equation}
Next, eliminating $\dot D_{l0\theta}$ from Eqs. \eqref{incremental-constitutive1}$_4$ and using \eqref{incremental-constitutive2}$_2$ and ultizing Eq. \eqref{incremental-solutions}, yields
\begin{equation}\label{line2}
U_\theta'=\frac{1}{r}\left[\frac{n(\gamma_{12}-\tau_{rr})}{\gamma_{12}}U_r+\frac{\gamma_{12}-\tau_{rr}}{\gamma_{12}}U_\theta+\frac{1}{\gamma_{12}}(r\Sigma_{r\theta})-\frac{n}{\gamma_{12}}\frac{\Gamma_{0122}}{K_{022}}\Phi\right].
\end{equation}
Similarly, eliminating $\dot D_{l0z}$ from Eqs. \eqref{incremental-constitutive1}$_5$ and using \eqref{incremental-constitutive2}$_3$ and ultizing Eq. \eqref{incremental-solutions}, yields
\begin{equation}\label{line3}
U_z'=\frac{1}{r}\left[\frac{kr(\gamma_{13}-\tau_{rr})}{\gamma_{13}}U_r+\frac{1}{\gamma_{13}}(r\Sigma_{rz})-\frac{kr}{\gamma_{13}}\frac{\Gamma_{0133}}{K_{033}}\Phi\right].
\end{equation}
Next, we substitute Eqs. \eqref{incremental-constitutive2}$_{2,3}$ and \eqref{incremental-solutions} into Eq. \eqref{incremental-Maxwell} and using Eqs. \eqref{line2} and \eqref{line3} to get the expression for $(r\Delta_r)'$, as follows
\begin{equation}\label{line4}
(r\Delta_r)'=\frac{1}{r}\left[\xi_1U_r-\frac{n\tau_{rr}}{\gamma_{12}}\frac{\Gamma_{0122}}{K_{022}}U_\theta+\frac{n}{\gamma_{12}}\frac{\Gamma_{0122}}{K_{022}}(r\Sigma_{r\theta})+\frac{kr}{\gamma_{13}}\frac{\Gamma_{0133}}{K_{033}}(r\Sigma_{rz})+\xi_2\Phi\right].
\end{equation}
These are the first four lines of the Stroh formulation.

Substituting Eqs. \eqref{incremental-constitutive1}$_{1,2,6,8}$, \eqref{incremental-constitutive2}$_{2,3}$ and \eqref{incremental-solutions} into Eq. \eqref{incremental-constitutive3}$_1$ and using Eqs. \eqref{line1}-\eqref{line3} results in 
\begin{multline}
(r\Sigma_{rr})' = \frac{1}{r}\left[ \kappa_{11}U_r+\kappa_{12}U_\theta+\kappa_{13}U_z-(\Gamma_{0111}-\Gamma_{0221})r\Delta_r+r\Sigma_{rr} \right. \\
\left.  -\frac{n(\gamma_{12}-\tau_{rr})}{\gamma_{12}}r\Sigma_{r\theta}-\frac{kr(\gamma_{13}-\tau_{rr})}{\gamma_{13}}r\Sigma_{rz}+\xi_1\Phi \right].
\end{multline}

Similarly, substituting Eqs. \eqref{incremental-constitutive1}$_{1,2,4,6,9}$, \eqref{incremental-constitutive2}$_{2}$ and \eqref{incremental-solutions} into Eq. \eqref{incremental-constitutive3}$_2$ and using Eqs. \eqref{line1} and \eqref{line2} gives
\begin{multline}
(r\Sigma_{r\theta})' = \frac{1}{r}\left[ \kappa_{12}U_r+\kappa_{22}U_\theta+\kappa_{23}U_z-n(\Gamma_{0111}-\Gamma_{0221})r\Delta_r+n(r\Sigma_{rr}) \right. \\
\left.  -\frac{\gamma_{12}-\tau_{rr}}{\gamma_{12}}r\Sigma_{r\theta}-\frac{n\tau_{rr}}{\gamma_{12}}\frac{\Gamma_{0122}}{K_{022}}\Phi\right].
\end{multline}

Then substituting Eqs. \eqref{incremental-constitutive1}$_{1,3,7}$ and \eqref{incremental-solutions} into Eq. \eqref{incremental-constitutive3}$_3$ and using Eq. \eqref{line1}, we obtain
\begin{equation}
(r\Sigma_{rz})' = \frac{1}{r}\left[ \kappa_{13}U_r+\kappa_{23}U_\theta+\kappa_{33}U_z-kr(\Gamma_{0111}-\Gamma_{0331})r\Delta_r+kr(r\Sigma_{rr})\right].
\end{equation}

Finally, from Eqs. \eqref{incremental-constitutive2}$_1$ and \eqref{incremental-solutions} and using Eq. \eqref{line1}, we have
\begin{equation}\label{line8}
\Phi'=\frac{1}{r}\left[(\Gamma_{0111}-\Gamma_{0221})U_r+n(\Gamma_{0111}-\Gamma_{0221})U_\theta+kr(\Gamma_{0111}-\Gamma_{0331})U_z-K_{011}r\Delta_r\right].
\end{equation}

Now we can write Eqs. \eqref{line1}-\eqref{line8} in the Stroh matrix form, as presented in Eq. \eqref{Stroh}.
\end{document}